\documentclass[fleqn,usenatbib]{mnras}
\usepackage{newtxtext}
\usepackage[T1]{fontenc}
\usepackage{ae,aecompl}
\usepackage{supertabular}
\usepackage{graphicx,longtable}	

\usepackage{amssymb,amsmath}

\def\sqiglt{\hbox{\rlap{\lower.55ex \hbox {$\sim$}}\kern-.05em \raise.4ex \hbox{$<$}\,}}
\def\sqiggt{\hbox{\rlap{\lower.55ex \hbox {$\sim$}}\kern-.05em \raise.4ex \hbox{$>$}\,}}
\def\til{\ensuremath{\sim\,}}

\def\chisq{\ensuremath{\chi^2}}

\newcommand{\tim}[1]{\ensuremath{\times 10^{#1}}}
\newcommand{\rev}[1]{\textcolor{black}{#1}}

\def\etal{et al.\ }

\def\xmm{\emph{XMM}}
\def\xmmn{\emph{XMM-Newton}}
\def\cms{\ensuremath{$cm$^{-2}}}

\def\swift{\emph{Swift}}
\def\rosat{\emph{ROSAT}}

\def\t0{\ensuremath{T_{0}}}

\def\P{\ensuremath{\mathcal{P}}}

\def\arcsec{\ensuremath{^{\prime\prime}}}
\def\arcmin{\ensuremath{^\prime}}
\def\nu{\ensuremath{\nu}}
\def\ergcms{\ensuremath{$erg cm$^{-2}$ s$^{-1}}}
\def\ecf{\ensuremath{$erg cm$^{-2}$ ct$^{-1}}}

\def\s{\ensuremath{\sigma}}

\title[LSXPS and transients]{A real-time transient detector and the Living Swift-XRT Point Source catalogue.}

\author[Evans et al.]{P.A. Evans$^1$\thanks{pae9@leicester.ac.uk},  K.L. Page$^1$, A.P. Beardmore$^1$, R.A.J. Eyles-Ferris$^1$,
\and J.P. Osborne$^1$, S. Campana$^2$, J.A. Kennea$^3$, S.B. Cenko$^{4,5}$
\\
$^1$School of Physics and Astronomy, University of Leicester, Leicester, LE1 7RH, UK \\
$^2$INAF, Osservatorio Astronomico di Brera, via E. Bianchi 46, 23807 Merate, Italy \\
$^3$Department of Astronomy and Astrophysics, Pennsylvania State University, 525 Davey Lab, University Park, PA 16802, USA \\
$^4$NASA Goddard Space Flight Center, 8800 Greenbelt Road, Greenbelt, DMD 20771, USA \\
$^5$Joint Space-Science Institute, University of Maryland, College Park, MD 20742, USA \\
}

\date{Accepted -- Received --}

\pagerange{\pageref{firstpage}--\pageref{lastpage}}
\pubyear{2022}

\begin{document}

\maketitle

\label{firstpage}

\begin{abstract} 

We present the Living \swift-XRT Point Source catalogue (LSXPS) and real-time transient detector.
\rev{This system allows us for the first time to carry out low-latency searches for new transient
X-ray events fainter than those available to the current generation of wide-field imagers, and report
their detection in near real-time. Previously, such events could only be found in delayed searches,
e.g.\ of archival data; our low-latency analysis now enables rapid and ongoing follow up of these
events, enabling the probing of timescales previously inaccessible. The LSXPS is, uniquely among X-ray catalogues,
updated in near real-time}, making this the first up-to-date record of the point
sources detected by a sensitive X-ray telescope: the \swift-X-ray Telescope (XRT). The associated
upper limit calculator likewise makes use of all available data allowing contemporary upper limits
to be rapidly produced on-demand. \rev{These facilities, which enable the low-latency transient system
are also fully available to the community, providing a powerful resource for time-domain and multi-messenger
astrophysics.}

\end{abstract}


\section{Introduction}
\label{sec:intro}

Serendipitous source catalogues have an important function in X-ray astronomy, helping us to
constrain the population of X-ray emitting objects, identify new phenomena not intentionally
targeted by observations and characterise the time-dependence of the X-ray sky. Such is their value
that a source catalogue is a standard product of most imaging X-ray telescopes, e.g. \rosat\
\citep{White94,Voges99,Boller16}; \xmmn; \citep{Saxton08,Watson09,Rosen16,Traulsen19};
\emph{Chandra} \citep{iEvans10} and the \emph{Neil Gehrels Swift Observatory}
\citep{Puccetti11,Delia13,Evans14,Evans20}. Some of these (e.g.\ \emph{Chandra} and the \xmm\
pointed catalogues) are narrow-field but deep; others (the \rosat\ all-sky survey [RASS], the \xmm\
slew survey, and the much-anticipated \emph{eROSITA} \citep{Predehl21} catalogues) are all-sky but
with less sensitivity.

The compilation of \swift\ X-ray telescope observations lies between these two extremes, covering
over 12\%\ of the sky, with typically more sensitivity than the all-sky surveys: in the 2SXPS
catalogue \cite{Evans20}, we reported a median 0.3--10~keV source flux of 4.7\tim{-14} \ergcms, not
much above the value from 3XMM-DR8 (2.2\tim{-14} \ergcms, 0.2--12 keV). Additionally, \swift\
typically observes a given object multiple times, giving a powerful insight into the variability of
the X-ray sky. This combination of sensitivity, sky area and variability information is likely to
remain unique until the multi-year \emph{eROSITA} catalogues are published.

\label{sec:swiftdata}

\begin{figure*}
  \begin{center}
  \includegraphics[width=16cm,angle=0]{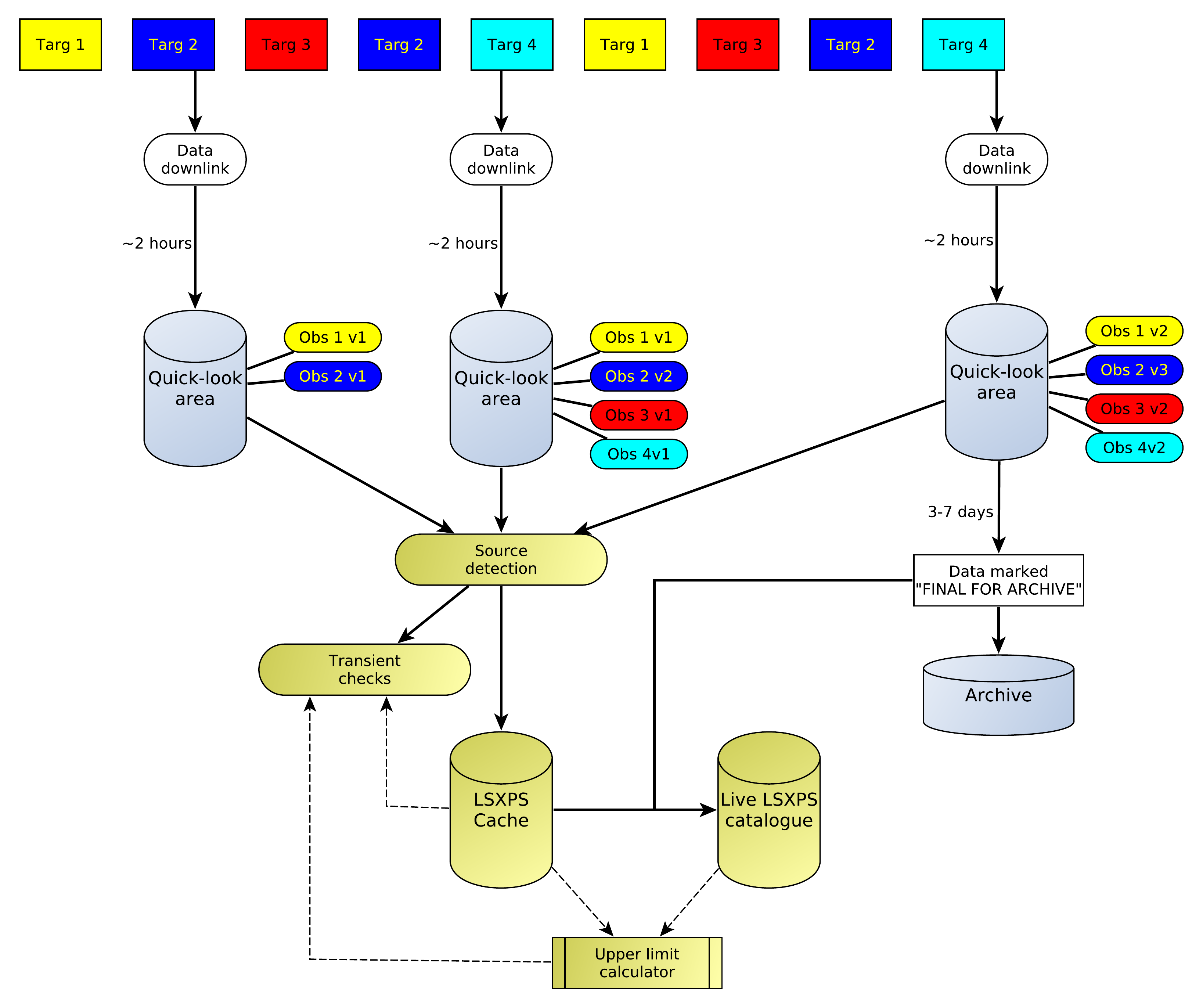}
  \end{center}
  \caption{A simplified schematic of the \swift\ data flow. The top row shows a timeline
  of \swift\ pointings in a given day, each box represents a pointing with \swift, multiple boxes with the same colour and label
  are multiple pointings at the same object (`snapshots'). Solid lines show data flow; dashed lines show
  data access.}
\label{fig:swiftObs}
\end{figure*}

However, there is a drawback to typical X-ray catalogues: they are always out of
date\footnote{Except those created after a mission has ended.}. Due to the time it takes to process
the data, compile the results, create a database and/or interface for catalogue access, and document
everything, the newest data are often at least a year old by the time the catalogue is released and
the observatory has collected more data in the meantime. For many purposes, this does not matter,
but with the recent growth of time-domain and multi-messenger astronomy, having an up-to-date
catalogue is increasingly important. For example, when observing a new gravitational wave event,
\swift\ regularly finds new X-ray sources, but to reliably determine which (if any) is related to
the triggering event requires prior knowledge of that sky location either to identify the source as
a known object, or to determine whether it is above historical upper limits \citep{Klingler19}.

Additionally, transients can be found serendipitously in X-rays; for example, the fast X-ray
transients found in \emph{Chandra} data \cite[e.g.][]{Jonker13,Bauer17,Quirola22}. These were found
by mining the data, but were already years old by the time of their discovery, preventing rapid
follow up \rev{and thus hampering the ability to expose the phenomenology of the transient, and the
underlying physics. Further, the variability timescales on which such transients can be studied are
limited to either the (hours to days) timescale of the discovery observation (e.g.\ the fast X-ray
transients, or nuclear quasi-periodic eruptions, \citealt{Miniutti19,Giustini20}), or potentially to
much longer timescales on which follow up can be carried out \citep[e.g.\
][]{Starling11,Strotjohann16}.} In contrast, \rev{low-latency announcement of transients allows
rapid, multi-wavelength follow up to probe the nature of the transient object, and enables ongoing
observations which can probe the source's behaviour on timescales of days to weeks/months. The
scientific value of such low-latency announcement and response is well demonstrated} by the case of
SN2008D. This supernova occurred in the same galaxy as another supernova (SN2007uy) which was being
regularly observed by \swift, and its sudden appearance in the X-rays was spotted by
\cite{Soderberg08}. This rapid discovery enabled targeted observations of what proved to be the
first X-ray detection of the shock breakout from a supernova.

\swift-XRT is an excellent tool for searching for new, serendipitous X-ray transients. Its
combination of good effective area and low background, combined with \swift's unique observing
stategy and the rapid availability of its data, allow us to look for transients in near real-time,
with sensitivity to much fainter transient events than from typical wide-field detectors such as the
RXTE All Sky Monitor \citep{Levine96}, or MAXI \citep{Matsuoka09}.

In this work we present the `Living' Swift-XRT Point Source (LSXPS) catalogue and real-time
transient detector. Software developed and running at the UK Swift Science Data Centre (UKSSDC) at
the University of Leicester is executed each time new \swift-XRT data are received. Those data are
searched for new transients, and added to the LSXPS catalogue, meaning it is kept up-to-date in
nearly real-time. The provision of the real-time catalogue is essential to the transient search: a
reference image is needed to determine whether a source just detected is already known in X-rays,
and if not, what upper limit can be placed on its historical flux. Because \swift\ often reobserves
locations on the sky (and a single XRT snapshot is usually more sensitive than the RASS or \xmm\
Slew Survey observations) it is commonly the case that the best reference image to use is a previous
\swift\ dataset, which is now always available in LSXPS.

\section{Data flow, from \swift\ to LSXPS}
\label{sec:dataflow}

  
In order to understand the `living' nature of LSXPS and the latencies associated with transient
detection, it is necessary to briefly introduce the structure and delivery of \swift\ data. A key
and novel feature of \swift's observing strategy is that it is always observing\footnote{More
accurately, nearly always: it does not observe while passing through the Southern Atlantic Anomaly
or slewing; due to its rapid slew speed, the latter is not a significant factor.}; when a target
goes into Earth eclipse, \swift\ immediately slews to the next target in its observing plan. Due to
its low-Earth orbit, the longest continuous observing window is \til2,700 s and for operational
reasons these are usually no longer than 1,800 s. This means that \swift\ both observes many targets
during a day and that the same target may be observed multiple times on different orbits, with
observations of other targets taking place in between. These individual pointings of a given target
are referred to as \emph{snapshots}, and (ordinarily\footnote{For technical reasons, sometimes an
observation may span multiple days, or there may be multiple observations in one day.}) all
snapshots of a given target within a single UT day are collected together on the ground into a
single \emph{observation}: the basic data unit used for archiving and data access. Each observation
is identified by its unique 11-digit obsID.

\swift\ typically has 8--12 ground-station contacts per day, and (again, unlike most missions) the
downlinked data are made available rapidly\footnote{Typically within about 2 hours of the downlink.}
via the quick-look sites in the USA, UK and Italy. These data are not necessarily `complete': not
all snapshots in a given observation may have occurred by the time of the downlink, or not all
observed data may have been downlinked during the pass. With each subsequent ground-station pass,
observations may be updated and the data on the quick-look site replaced with a revised (more
complete) version; to track this, metadata indicates how many times the observation has been
processed. Eventually, each observation will be marked as `complete' by the Swift Data Center (SDC),
normally 3--7 days after the data were collected. The quick-look sites will label the data as `FINAL
FOR ARCHIVE', and shortly thereafter these data are moved from the quick-look area to the archive.

A simplified pictoral representation of this is given in Fig.~\ref{fig:swiftObs}. The coloured boxes on the top
row represent consecutive pointings; ground station passes are shown with the state of the quicklook site after each
pass. 

\subsection{Adding data to LSXPS}
\label{sec:LSXPSdata}

Quick-look XRT data are by definition incomplete, and can also occasionally be `incorrect', in the
sense that the incomplete housekeeping data can result in the pipeline processing inadequately
filtering the data. For this reason, observations are not added to the actual LSXPS catalogue until they
are marked as `complete' by the SDC. Analysis of these incomplete observations is, however,
extremely important for transient detection; not just for the obvious reason of finding transients
as early as possible, but also to provide the most complete possible set of reference images for
future searches\footnote{For example, imagine \swift\ observes the field of a new GRB and finds no
new serendipitous transient. The GRB is re-observed the following day -- before the first
observation is marked as complete. If X-ray sources are found in this second observation, the
analysis of the first observation is vital to determine if any of these sources are new
transients.}. 

For this reason, observations are analysed each time data are received by the UKSSDC quick-look
site, as shown in Fig.~\ref{fig:swiftObs}. The receipt of quick-look data triggers the LSXPS source
detection, and the results are stored in a cache area which it not made publicly available.
Transient checks (Section~\ref{sec:transCheck}) are immediately carried out for each detected object and the XRT
team are notified of any transient canidates, but none of the sources found in the observation, nor
the observation itself, are added to the public catalogue (although the observation is available to
the upper-limit server, Section~\ref{sec:access}). The transient checks for subsequent observations
can make of use these cached data. When a revised version of an observation is received on the UK
quick-look site, it is analysed and the cached results for the observation are replaced with those
from this new analysis. In order to track this evolving situation, for LSXPS we index data not by
the ObsID, but by a new property, the \emph{DatasetID}. This is simply an incremental integer number
starting from 1, and each new version of an observation is given a new DatasetID, so `Obs 1 v1' and `Obs \rev{1} v2'
in Fig.~\ref{fig:swiftObs} will have different DatasetIDs\footnote{DatasetIDs
are set simply by the order in which data are processed; thus two different version of the same
observation will not normally have consecutive DatasetIDs.}. Once the observation is marked as
complete, the cached analysis is merged into the catalogue (see Section~\ref{sec:addSources}).

\subsection{Targets, observations; stacked image definition and evolution}
\label{sec:stacks}

\swift\ observations are planned in terms of \emph{targets}, where a target, with a unique targetID,
corresponds to a pointing location on the sky. Thus multiple observations of the same target will
share a targetID; indeed, the 11-digit obsID is just the 8-digit targetID followed by a 3-digit
number tracking how many times the target has been observed. Occasionally, either due to a human error or
an inaccurate slew, two observations with the same targetID have significantly different
pointings; if observations with the same targetID are more than 6\arcmin\ offset from each other
then within LSXPS we define new targetIDs for each of these pointings, with observations being
assigned to the targets as appopriate.

As with the previous SXPS catalogues, as well as analysing individual observations, we also create
\emph{stacked images}, where all observations of the same portion of the sky are combined to
maximise the exposure\footnote{We slightly increased the maximum size of a single stacked image to
2,700 pixels (106\arcmin) to a side for LSXPS.}. A stacked image is defined as a specific set of
overlapping targets (or a single target, with multiple observations), and the population of stacked
images is created such as to produce the minimum number of stacks while ensuring that every
observation and every overlap between observations is included in at least one stack. As in the
previous catalogues, stacked images are assigned fake obsIDs, incremental numbers begining with
$10^{10}$, and the term \emph{dataset} is a generic term covering individual observations and
stacked images. 

In the previous catalogues, stacked images were statically defined, but this is clearly not possible
for a living catalogue: not only are new stacks being constantly added, but the existing stacks can
be modified. This can happen in two ways. First, a new observation may be taken of a target in a
stack. In this case, the LSXPS processing does not change the `ObsID' of the stack, but does assign
it a new DatasetID (how many times this stack has been analysed is also recorded); this is
deliberately analogous to what happens to single observations when they are updated with more data.
Second, a new target may be observed which overlaps an existing stack. In this case, the definitions
of the stacks are updated. It is impossible to say a-priori what the outcome of this will be. In the
simplest case a new stack is defined which is the same as an old stack but with the new target
added; however, the effect may be that the addition of the new target causes a stack to grow beyond
the 2,700-pixel limit in which case multiple new stacks may be defined, or the new target may bridge
a gap between two existing targets and so be able to merge all of their data into a single (new)
stack. In either case, stacks that are superseded are marked as such, and are not analysed when new
data are received. The handling of superseded stacks is discussed further in
Section~\ref{sec:oldStacks}.

\subsubsection{Analysing stacks}
\label{sec:whenRunStack}

Stacks can take much longer to analyse than single observations -- the runtime scales approximately
with the number of snapshots -- and in the most extreme cases stacked image analysis can take many
weeks to run and its temporary disk usage can near a terabyte. It is thus impratical to simply run
stacked analyses whenever new data are received, as is done for observations. Further, there is less
need for this: as a general rule, each new observation of a stack adds only a (ever-decreasing)
fractional increase in exposure time, and we do not search stacks for transients
(Section~\ref{sec:transCheck}) -- their primary function for transients being the provision of
historical upper limits. Therefore, stacked image processing is instead managed by a {\sc cron} job. This checks
how many stacked image analyses are in progress, and if it is above some threshold\footnote{Both
this threshold, and the frequency with which the {\sc cron} job runs are tunable parameters,
allowing us to change things as our compute capactiy varies.} it will terminate without submitting
any new jobs. Otherwise, it identifies $n$ observations (where $n$ is again a tunable parameter)
which were marked as complete at least 24 hours ago and have not yet been subjected to stacked analysis, finds
all stacks that these observations contribute to, and triggers the analysis of those stacks. This analysis
will include all observations in the stack which are marked as complete.

Long-running, high disk-usage processes are still an issue for this approach, since they can
fill up all of the available stack-processing `slots', causing the {\sc cron} job to exit without
scheduling any new fields for days or even weeks. To avoid this, every time a stacked image is analysed
the runtime is recorded in a database. Stacks which took more than 24 hours to run (or which supersede a stack
which took more than 24 hours) are not handled by the routine {\sc cron} job just described, but are instead
flagged in the database, and analysed at a lower cadence (loosely connected to the runtime) in a separate queue,
as resources allow. By definition, these slow-running fields already contain a lot of data, therefore the
incremental changes are small and the impact of this low-cadence update is minimal.

\section{Catalogue construction}
\label{sec:build}

The basic processing of a dataset for LSXPS remains almost the same as in 2SXPS \citep{Evans20} and
is not described in detail here: a brief overview is given in Appendix~A and the interested
reader is encouraged to read the 2SXPS paper for full details; all we will do here is to restate
the four energy bands used in LSXPS:

\begin{itemize}
  \item {\bf \em Total}: 0.3 -- 10 keV
  \item {\bf \em Soft}: 0.3 -- 1 keV
  \item {\bf \em Medium}: 1 -- 2 keV
  \item {\bf \em Hard}: 2 -- 10 keV
\end{itemize}

\noindent \rev{and reproduce details of the source classification system. Sources are allocated a
detection flag based on their statistical properties, these can have values of \emph{Good},
\emph{Reasonable} and \emph{Poor} (in the database tables, these are numerical values of 0, 1, 2
respectively). The thresholds for these parameters were set such that the fraction of spurious
sources in a given sample of objects is 0.003, 0.01 and 0.1 if that sample includes \emph{Good},
\emph{Good} and \emph{Reasonable} or all sources respectively; individually the classifications
correspond roughly to 3-, 2- and 1-\s\ significance. The spurious source rate will be considerably
higher in the presence of artifacts such as stray light, or an extended/diffuse source. For this
reason, objects which are affected by such things (identified either automatically or by the manual
screening, Section~\ref{sec:screen}) have extra warning flags set; on the website this is shown as
descriptive text; in the database these flags are bitwise values in the detection flag, as shown in
Table~\ref{tab:detwarn}. Datasets likewise have flags assigned to them to warn if there is a
potential issue affecting the dataset as it is possible (but unlikely) that these issues could alter
the spurious detection fraction even among objects in that field with none of their warning flags
set. This flag is also a binary flag, detailed in Table~\ref{tab:fieldflag}.}

\begin{table}
  \begin{center}
  \caption{The warning flag bits in the detection flag; taken from Evans \etal(2020), table 4.}
  \label{tab:detwarn}
  \begin{tabular}{ccc}
  \hline
Bit & Value & Meaning \\
  \hline
2   &  4  & Source is within the extent of a known \\
    &  &  extended source. \\
3   &  8  & Source likely a badly-fitted piled-up source. \\
4   &  16  & Position matched area flagged \\
    &     & by manual screening. \\
  \hline
  \end{tabular}
  \end{center}
\end{table}

\begin{table}
  \begin{center}
  \caption{The definition of the dataset warning flag, taken from Evans \etal(2020), table 5.}
  \label{tab:fieldflag}
  \begin{tabular}{ccc}
  \hline
Bit & Value & Meaning \\
  \hline
0   &  1  & Stray light was present, and fitted. \\
1   &  2  & Diffuse emission identified. \\
2   &  4  & Stray light badly/not fitted. \\
3   &  8  & Bright source fitting issues$^1$ \\
  \hline
  \end{tabular}
  \end{center}
  \begin{flushleft}
    $^1$ i.e.\ the field contained a source that was heavily piled up in one band, but not fitted as such
    in another band. See section~3.7 of \cite{Evans20} for details.
  \end{flushleft}
\end{table}

There is one small but important extension to the 2SXPS algorithm for LSXPS for individual
observations: the search for new transients, which is described in Section~\ref{sec:transCheck}.

LSXPS consists of all XRT data with at least 100~s of Photon Counting (PC) mode exposure
remaining after the standard {\sc xrtpipeline} processing at the UKSSDC, and after the removal of
times affected by bright Earth or unreliable pointing (Appendix~A). These data are
split into snapshots, only snapshots with at least 50~s of exposure are accepted\footnote{This
can result in LSXPS datasets with only 50~s of exposure, for example if the original 100~s of usable
data was spread over multiple snapshots, only one of which exceeded 50~s in duration. This was true
of the previous SXPS catalogues but has not been explicitly stated before.}. All XRT data meeting
these criteria were reprocessed with {\sc heasoft} v6.29, and the XRT CALDB v20210915\footnote{At the time of writing,
all LSXPS data use these software and CALDB versions; however, they will doubtless be updated again
in the future at which point only new data in LSXPS will use the newer software and calibration: we
have no plans to periodically rebuild LSXPS from scratch following such updates.} and then the
catalogue creation software was run on all historical data (we did not copy the results from the
earlier catalogues). This processing `caught up' with the archive data towards the end of 2022
March, and after a series of short `catch-up' runs, the `live' processing software to handle new
quick-look data as described above was enabled on 2022 April 1\footnote{I did consider delaying by
one day\ldots}. The analysis of the stacked images built from historical data was also enabled at
this point, with the queue of images to process automatically updated by the LSXPS processing as new
data were received and so new stacks defined. Once the stacked image processing was
up-to-date\footnote{Or \emph{almost} up to date. Various technical challenges were encountered
relating to certain stacks with very long ($>100$ hour) runtimes, almost all of which cover the
Galactic centre. Analysis of these images was deferred until the backlog was cleared, and is now --
slowly -- catching up.}, the {\sc cron} job described in Section~\ref{sec:whenRunStack} was
activated.

While the source detection algorithm is almost unchanged from 2SXPS, the aggregation of the results
to form the final catalogue required significant modification for LSXPS. The method by which the
various detections are rationalised to a unique source list is unchanged (see section~3 of
\citealt{Evans20}), but for 1/2SXPS this step was done once, after all datasets had been processed
and the list of detected objects was final. For the dynamic LSXPS catalogue this is a continuous
process, and takes place when the dataset is added to the already-existing catalogue. For observations this is
done when the data are marked as complete; for stacks this step is carried out immediately after
the dataset processing is completed.

There are four steps involved in adding a dataset to LSXPS, which are:

\begin{enumerate}
\item Manual screening of the field$^\dagger$.
\item Creating a new entry in the datasets table.
\item Incorporating the objects detected in the dataset into LSXPS.
\item Handling older versions of a stacked image$^\dagger$.
\end{enumerate}

\noindent Items marked with $^\dagger$ may not take place for all fields. With the exception of
item (ii) (which is just a database operation), these operations are described in the following subsections.

\subsection{Manual screening of fields}
\label{sec:screen}

In order to make the quality of LSXPS as high as possible, the analysis software identifies datasets
in which the results may be unreliable and flags these for human inspection, which must be completed
before the dataset can enter the catalogue. There are two criteria that trigger this flagging.

The first of these relates to stray light and is not applicable to stacked images. X-rays from a
source \til30--80\arcmin\ from the XRT boresight (and so out of the field of view), can be diverted
onto the XRT detector via a single reflection off the hyperbolic mirror surface; if the source is
bright enought this can cause a pattern of concentric rings to appear on the XRT detector, which can
give rise to spurious detections. We developed a technique to identify and fit this emission for
2SXPS (see \citealt{Evans20}, appendix A), but this is not infallible, especially if there are real
X-ray sources in or near the stray light. Such errors are immediately obvious to the human eye, so
any field for which the automated system deemed stray light to be present is marked for human
checking. Once this dataset is marked as `final', the XRT team are notified and directed to a web
interface which allows them to compare the image with the fitted background model (which includes
the stray light), and to either accept the model, mark the stray light as not being present (i.e. a
spurious fit), or mark it as badly fitted and provide a better input position for the source of
stray light. In the first case, the dataset is accepted into LSXPS. In the latter two cases, the
dataset is reanalysed (reusing its existing DatasetID, as the underlying data have not changed),
i.e. with stray light modelling disabled or with the new input parameters as appropriate. As soon as
this reanalysis is complete, the field is again marked for screening. In the vast majority of cases
it is then accepted into the catalogue. In a small number of cases, the catalogue software is not
able to produce an adequate model for the stray light. In this case the XRT team will define an
artifact (see below) covering the parts of the image affected by stray light, causing all detections
therein to be flagged. As described in appendix A of \cite{Evans20}, stray light is not fully fitted
to stacked images; instead the results of the fits to the observations making up the stacked image
are taken and applied to the stacks, with only the normalisation allowed to vary. For this reason,
if the stray light fitting for a stack is poor, this must be handled by the creation of artifacts,
as below.

The second reason for human verification of a dataset is if median inter-source distance in the
dataset is less than 80\arcsec. This can simply indicate a full or crowded field, but it can also
indicate problems such as badly- or unfitted stray light, instrumental artifacts such as residual
contamination by bright Earth or hot pixels/columns/rows in the detector, or an area of diffuse
emission which is not handled by our point-source detection system and so causes spurious
detections. Such fields are flagged and the XRT team again notified and directed to a website where
they can investigate the image. If there is evidence of a problem, they will define a circular or
elliptical region, or a set of regions, and all of the objects inside these and in this specific
dataset have a warning flag applied to them. If the `problem' is astrophysical in origin (as opposed
to being of instrumental origin),  e.g. an extended source, then the team member will mark these
artifacts for propagation, which will mean that they are automatically applied to all future
observations of these location on the sky. This helps to reduce the number of fields needing
screening, since fields are only marked for inspection if the median distance between sources
\emph{that have not already had their warning flag set} is below 80\arcsec.

For datasets corresponding to observations in the 2SXPS catalogue, the results of the human
screening of those data in that catalogue were automatically applied where possible. That is, if the
automatic stray light model closely corresponded to that accepted for 2SXPS, it was accepted for
LSXPS. Artifacts identified in 2SXPS were automatically applied to their corresponding LSXPS fields,
and those fields were therefore only flagged for LSXPS screening if the unflagged sources has a low
median separation.

\subsection{Adding sources to LSXPS}
\label{sec:addSources}

When a dataset is added to LSXPS, the objects detected in that dataset are compared with the sources
already in LSXPS. An object is deemed to match a source in LSXPS if their positions -- including
astrometric uncertainty -- agree at the 5-\s level\footnote{More correctly, at the probability
associated with a Gaussian 5-\s\ confidence. Since the radial position errors should follow a
Rayleigh distribution, this level was determined based on Rayleigh, not Gaussian, statistics.}.
There are three potential results for each new object:

\begin{enumerate}
  \item The object has no counterpart in LSXPS.
  \item The object is an additional detection of a source in LSXPS.
  \item The object matches an LSXPS source, but improves the position of the LSXPS source (i.e.\ its
  position uncertainty is smaller than that currently in LSXPS).
\end{enumerate}

\noindent The first two are straightforward: in case (i) the object is a new source which is added to LSXPS;
in case (ii) the products and average properties of the existing source are updated to include the results
from the new dataset. 

Case (iii) is more complex as there are several possible permutations. In the simplest scenario, the
new position is identical to the LSXPS position\footnote{We define `identical' as meaning, `does not
change the IAU-format name of the object', where the IAU format is ``LSXPS JHHMMSS.S$\pm$ddmmss''.},
but the uncertainty is smaller. Here, the catalogue entry is updated to use this smaller error and
the processing continues as per case (ii), above. If, however, the new position is different from the
LSXPS position, the situation is more complex as the association between individual (per-dataset)
detections and LSXPS sources may change. For example, if there are several X-ray sources close
together on the sky, then moving one of these sources 1\arcsec\ to the East could result in some
detections previously assumed to be this source now being better associated with a different object;
or indeed the inverse. \rev{On the other hand, if a source position is revised it may now be the case
that some of the detections, which had been attributed to that source, are not consistent with the new
position, nor are they consistent with that of any existing source; i.e.\ the original `source' is now
found to be two distinct sources, which of course each need a record in LSXPS.}

In order to identify and handle all of these cases, when a dataset is added to the catalogue we
select every LSXPS source that lies within the field of view of the new dataset, and every
individual detection of those sources; then we find any other sources within 5-\s\ of those
detections, and all detection of these; and so on recursively until no extra objects are found. This
is analogous to picking up a paperclip with a magnet and so also retrieving all connected paperclips
until a gap (in this case, a 5-\s\ gap) is found, and essentially ensures that we collect every
single detection in any dataset and band that could potentially be affected by the new dataset. We
then apply the source matching algorithm from 2SXPS (\citealt{Evans20} section~3.7) to this subset
of detections. This yields the list of unique sources and which detections are associated with each,
and the catalogue is revised accordingly. If a source moves such that its name changes, a new entry
is created in the catalogue and the old entry deleted, but a record of this is kept. Records are
also kept of which sources are deconvolved into multiple sources, and which individual detections
are reassigned from one source to another. This book-keeping is extremely important for the
usability of a dynamic catalogue such as LSXPS: if someone studies a particular source, perhaps
announcing it to collaborators or the wider community, and then the source disappears without a
trace this would be very unhelpful! The user interface (Section~\ref{sec:access}) ensures that if a
user requests a `superseded' source they will either be redirected to the new entry which has
replaced this source, or if necessary presented with a list of the sources into which it has been
deconvolved.

\subsubsection{Source products and updates}
\label{sec:sourceProd}

For each source, a series of products and measurements are created, and these are the same as in
2SXPS (see \citealt{Evans20}, section~4). In short, these comprise mean count-rates in each energy
band; two time-averaged hardness ratios; and light curves of all energy bands and hardness ratios, with one bin
per observation, and one bin per snapshot. Unlike some catalogues, these light curves and
count-rates are fully corrected for effects such as pile up, vignetting, dead columns on the
detector etc. and so can be used with no further corections needed. A measurement of variability
using the Pearson's \chisq\ is calculated for each time series. The two hardness ratios are defined
as:

\begin{equation}
  \label{eq:HR1}
  {\rm HR1} = (M-S)/(M+S)
\end{equation}

\begin{equation}
  \label{eq:HR2}
  {\rm HR2} = (H-M)/(H+M)
\end{equation}
  
\noindent where $S, M, H$ refer to the background-subtracted count-rates in the soft,
medium and hard bands respectively. 

Spectral and flux information is also provided for each source, with up to three methods employed
per source, depending on the characteristics. Two spectral models are used: an absorbed power law,
and an absorbed optically thin plasma model (APEC; \citealt{Smith01}); absorption was calculated
using the TBABS model \citep{Wilms00}. Flux conversions and (where appropriate) spectral
properties were derived using {\sc xspec}. For every source, fluxes were deduced using fixed
spectral components: a power law with photon index 1.7 and an APEC with a plasma temperature of 1
keV; the absorption column was set to the Galactic value along the line of sight to the source, from
\cite{Willingale13}. These values are available for every source but are the least accurate since
they simply assume the spectral shape, regardless of the data. Where possible, the spectral parameters
and flux are also inferred from the hardness ratios, using look-up tables linking (HR1,HR2) to the model
parameters -- for full details see \cite{Evans20} section~4.1 and \cite{Evans14} section~4.2.
For sources with at least 50 net counts in the total band, and a detection flag of \emph{Good} or \emph{Reasonable} (Section~\ref{sec:contents}) with
no stray light or diffuse source warning set, we also extract and fit a spectrum. This is done using the tools of \cite{Evans09},
and we fit both absorbed power-law and APEC models.

As LSXPS is a living catalogue, these source properties are not static. Every time a new dataset is added
to the catalogue, all sources that lie within the field of view of that dataset (regardless of whether or not they
were detected in it) have their products updated. 

\subsection{Handling older versions of a stacked image}
\label{sec:oldStacks}


When a new version of a stacked image is added to the catalogue it is preferable to remove the old
version, rather than keeping large numbers of obsolete datasets. If every source detected in the old
version is also detected either by the new version, or by another dataset, then there is no reason
to keep the old version of the stacked image. Its data are removed from disk and from most of the
database tables\footnote{A record of all such deleted datasets is still maintained but this is not
part of the public catalogue.}. It can happen, however, that a source is detected in a stacked image,
but not in the updated version of that stack -- especially for faint and variable or transient sources. If the
old version of a stacked image contains the only detection of a source in LSXPS, that stacked image
will not be removed from the catalogue. Instead, the source is flagged internally as an `orphan', and the 
dataset is marked as being an `obsolete stack'. Should an orphan source ever be detected in a new dataset (single
observation or stacked image), its orphan status is removed; should this happen to all orphans in an obsolete stack,
that stacked image is then removed. The same process is followed when a new stacked image is analysed which supersedes a stack
currently in the catalogue; that is, the new stack has a different definition (and obsID), but
contains all of the targets in an old stack (see Section~\ref{sec:stacks}).

\subsection{Static catalogue releases and data reprocessing}
\label{sec:static}

The introduction of a living catalogue, with near real-time updates, does not remove the need for
`static' catalogues, for projects which require a stable and unchanging reference such as where the
reproducibility of a project is crucial. It is therefore our intention to issue frozen releases of
the catalogue (3SXPS, 4SXPS, etc.) at regular intervals, likely every 1 or 2 years. These will
essentially be `snapshots' of LSXPS, but will be full decoupled from LSXPS, so completely insulated
from updates. Because of the timescales on which stacked image analysis runs, it may not be possible to fully
`synchronise' the contents of these releases; that is, to ensure that the stacks in a static release contain
all of the observations in the release, and only those observations. We do not see this as a major issue, as the
key point of these releases is that their contents can be quantified and are unchanging, and this will be ensured.
The first static release, 3SXPS, is envisioned for 2023.

Before building LSXPS, we reprocessed all historical XRT data with what was, at the time, the most
recent {\sc heasoft} and XRT calibration releases (Section~\ref{sec:build}). However, these are
regularly updated\footnote{Indeed, {\sc heasoft} 6.30 was released during the processing.} and
calibration in particular is, by its nature, always retrospective. While the UKSSDC does
periodically reprocess the entire XRT archive following such updates, we are not planning to rebuild
the LSXPS catalogue following such processing. The CALDB and {\sc heasoft} versions used in the live LSXPS analysis will be
updated, but the data already in the catalogue will not be revised. The reason for this is that such
a revision would essentially involve completely rerunning the LSXPS analysis: a process that takes
around 6 months and monopolises compute time, and so clearly cannot be done routinely! It is not
anticipated that any software or calibration revisions will have a major impact on the catalogue
contents\footnote{In the unlikely event that something changes that is impactful, we will
investigate how best to respond, and in such cases a complete rebuild of the catalogue cannot be
ruled out.}. Changes in the XRT gain could potentially have a (very) small impact on the hardness
ratios and spectral properties of sources; however, the catalogue web pages and the API (Section~\ref{sec:access})
are fully integrated with the UKSSDC software to provide high-quality analysis of point sources
\citep{Evans09}, making it easy to generate custom spectral products for sources of interest, making use of the
latest software and calibration, and using the archive data locally reprocessed with these at the UKSSDC.

\section{Transient detection and classification}
\label{sec:transCheck}

We define a serendipitous transient as an object which is uncatalogued in X-rays, has a flux above
historical upper limits and was not the target of the \swift\ observation. In order to minimise the
delay in detecting and announcing these transients we do not wait for an observation to be marked as
complete before carrying out transient checks, but rather carry out these checks immediately after each
analysis of an observation.

The transient checks applied to each detected object can be broken into the following steps:

\begin{enumerate}
  \item determine whether the object is a known X-ray source
  \item determine whether it is above historical upper limits
  \item carry out basic analysis of the object
  \item classification of the object
\end{enumerate}

\noindent and these steps are described in the following sections. At each step the process may
terminate; e.g.\ if the source is a known X-ray emitter, then steps (ii) onwards are not carried
out. 

During the bulk processing of historical data, the transient checks were carried out, but notifications
of new transients (see below) were not produced and the candidate transients were not added to the main
LSXPS transients database but stored separately. This is because the events are, by definition, old and so 
not pressing, and may have been announced or published already. The primary purpose of enabling transient
checking and analysis during the historical process was to test and debug the software and tune
the automated classification process (Section~\ref{sec:classify}); however, we may release the database
of these historical transients or some analysis thereof at a future date.

\subsection{Is the object an known X-ray source?}
\label{sec:known}

We deem an object to be uncatalogued in X-rays, if it does not correspond to a source already in
LSXPS (or in the cache\footnote{Excluding sources found in an earlier version of the observation
being analysed.}), and is inconsistent at the 5-\s\ level with any object in the X-ray Master
catalogue provided by {\sc
heasarc}\footnote{\url{https://heasarc.gsfc.nasa.gov/W3Browse/all/xray.html}.}, CSC 2.0 \citep{iEvans20} or the live XRT GRB catalogue
\citep{Evans09}\footnote{\url{https://www.swift.ac.uk/xrt\_live\_cat}}. This set of reference
catalogues may be updated in future (for example, with the addition of \emph{eROSITA} catalogues),
an up-to-date list will be maintained on the LSXPS website.

\subsection{Comparison with historical upper limits}
\label{sec:transUL}

We compare the flux at discovery with upper limits obtained from \swift\ itself and also \xmmn\
and \rosat\footnote{This list may also be revised in future, especially to include \emph{eROSITA}; the
LSXPS website will maintain an up-to-date list}.
\xmm\ has a similar band-pass to \swift-XRT and we can use the total (0.3--10 keV) LSXPS band for
this analysis. \rosat, however, has a much softer band-pass, 0.1--2.4 keV, thus for comparison with \rosat\
we use the combination of the soft and medium (0.3--1, 1--2 keV) LSXPS energy bands, hereafter the `SM' band.

We create light curves with one bin per snapshot (i.e.\ \swift\ orbit) in the total, soft and medium
LSXPS energy bands, summing the latter two to give the SM-band light curve. The peak per-snapshot
count-rate is then identified from these light curves, defined as the bin with the highest 1-\s\
lower-limit\footnote{This definition is robust against datapoints with very large errors; a
count-rate of $0.5\pm0.05$ is deemed higher than a rate of $2\pm1.9$.}. We also calculate the mean
count-rate of the object across the observation in the total and SM bands; if this has a higher
1-\s\ lower-limit this is instead taken as the peak rate. To enable us to compare upper limits in
the next step, we also calculate the ratio, $R$, of the peak count-rates in the total and SM bands.

3-\s\ upper limits are then obtained from LSXPS (including cached datasets and stacked images, see
Section~\ref{sec:access}), \xmmn\ and \rosat; the latter two instruments' limits are obtained using
the HILIGT service\footnote{\url{xmmuls.esac.esa.int/hiligt/}.} \citep{Saxton22,Konig22}. Of these,
only the RASS is an all-sky survey thus not all upper limits are always available. Upper limits are
created in count-rate units using the native energy band of the instrument (0.2--12 keV for \xmm,
0.2--2 keV for \rosat\ and 0.3--10 keV for LSXPS); for \xmm\ limits from both pointed and slew data
are requested. These limits are then converted to XRT count rates, using fixed conversion factors
determined using {\sc pimms} and assuming a power-law spectrum with spectral index 1.7 and an
absorbing column of $10^{21}$ \cms; for \xmm\ the conversions were calculated for each instrument
and filter and so the appropriate one is selected. \xmm\ upper limits are converted to total-band
rates, and \rosat\ limits to SM band rates. The deepest of these upper limits is then taken as the
reference value to use. Since the \rosat\ limit is in a different energy band from the others, for
the comparison we multiply it by the ratio $R$ defined above which helps to account for the spectral
shape of the object, which is unknown at this point. Having determined the deepest upper limit, the
peak count-rate from the appropriate band is compared with this limit, and if it is at least 1-\s\
above it, the object is flagged as a candidate transient\footnote{Note that the \rosat\ limit extrapolated
to the 0.3--10 keV band is \emph{only} used for comparing with the upper limits to determine which is deepest.
If \rosat\ is selected by this test, it is the upper limit in the SM band which is used, and compared with the
peak SM-band count rate.}.

During the historical processing, we found that all transient candidates that had a detection flag
of \emph{Poor} or one of the warning flags set (see Appendix~A) were spurious detections, and indeed a
significant fraction of alerts came from such objects (this is expected: a real source with a
\emph{Poor} detection is unlikely to be bright enough to pass the above tests, and while transients
occuring within a stray light ring or known extended object are not impossible, they will be rare,
whereas spurious detections with high apparent count-rates are much more common). Accordingly,
such candidates are automatically classified as \emph{Spurious} (Section~\ref{sec:classify})
and the analysis (Section~\ref{sec:transAn}) is not carried out and the XRT team is not alerted
to such events. They are still posted to the team-only website discussed below, and they can be
`upgraded' if a team member studying the image believes them to be real, but they are not routinely analysed.

\subsection{Basic transient analysis}
\label{sec:transAn}

For each candidate transient identified by the previous steps, custom data products are built using
the tools that power the UKSSDC on-demand analysis
services\footnote{\url{https://www.swift.ac.uk/user\_objects}.} \citep{Evans09}. A light curve is
built using all available XRT data, including those taken before the transient discovery (if
available); this can result in evidence of historical emission from the source even though it has
not been previously detected, since forced-photometry at the location of a known source is more
sensitive than a blind search\footnote{In each bin, if the 3-\s\ lower-limit on the count-rate is
non-zero, a `detection' comprising a count-rate and 1-\s\ errors is produced; otherwise a 3-\s\
upper limit is created}. This is binned in three ways: with fixed numbers of counts per bin, with
one bin per snapshot, and with one bin per observation. A spectrum is also created using only the
discovery observation; this is partly to maximise the S/N, and also to determine the spectral state
specifically at the moment of outburst. This spectrum is automatically fitted with an absorbed
power-law model, and if the automatic fit is succesful and the reference upper limit
(Section~\ref{sec:transUL}) is not from LSXPS, then the upper limit is revised, the conversion
from \xmm\ or \rosat\ to XRT being recalculated with this fitted model, using {\sc pimms} and
the best-fitting model parameters.

These data products and the revised upper limit are posted to a website accessible only to the
XRT-team, who are then notified by email of the potential transient and asked to confirm its status and
classify it. This happens even if the redetermined upper limit is no longer below the peak
count-rate, in case the spectral fit was poor; however the notification email does indicate if this
was the case.

\subsection{Transient classification}
\label{sec:classify}

The final step of transient analysis requires human checks to filter out objects which are
definitely not serendipitous transients, and to further classify those which are. This step is
currently semi-automated: the analysis system tries to suggest a classifcation, but a human is still 
required to verify or change this before the transient is made public. 

Objects may be deemed not to be serendipitous transients for the following reasons:

\begin{itemize}
  
  \item {\bf Spurious detection}: the `transient' may be an obviously spurious detection. Many of these cases
  are identified automatically as noted above but some may not be -- for example if the X-ray `source' is actually
  an artifact of optical loading, the automated system will (usually) warn that this is possible, but
  will not automatically flag the source as spurious.
  
  \item {\bf Targeted transient}: the object is a transient, but is not serendipitous, it is the
  object \swift\ was observing (e.g. as a target-of-opportunity observation of a transient
  discovered elsewhere). These cases can usually be flagged automatically by the analysis code,
  which uses the {\sc ObsQuery} class in the {\sc swifttools Python}
  package\footnote{\url{https://www.swift.ac.uk/API}.} to determine what object was
  the target of the observation. In cases where the target observation had a large position
  uncertainty, the automatic flagging is less reliable.

  \item {\bf It is not a transient}: analysis of the historical light curve may show a series of previous bins
  in which the source flux can be constrained, and is consistent with that at the time of discovery. In this
  case the source is clearly not transient, but as a result of the specifics of the current observation
  happens to have been blindly detected for the first time. Alternatively, the spectral fit to the transient
  data may cause the historical upper limit to be above the peak flux; if the XRT team member confirms that
  this fit is reliable, the transient nature of the source is retracted.

\end{itemize}

There is an additional category that is not listed above, a \emph{proprietary transient}. This is an
object which passes the above tests and is a real transient, but the parent observation was part of
a programme designed to monitor a specific region to look for new transients. In this case whether
the transient should be deemed `serendipitous' is open to interpretation, but it is clearly
inappropriate for the \swift\ team to routinely and rapidly announce the primary science of
someone's approved observing programme, and before they have even had time to examine the data! In
these cases, the \swift\ team will endevaour to contact the observation PI and encourage them to
announce the transient, and the object will not be posted to the public transients catalogue or
webpage without their consent.

Items which are not caught by the above are still not necessarily transients, and are manually classified into
one of the following groups, collectively referred to as `interesting classes'.

\begin{itemize}
  
  \item {\bf Outburst}: If the light curve shows historical detections of the source (from the forced photometry) but at
  a flux level below that of the new observation, the source is likely an outburst rather than a transient.
  
  \item {\bf Low significance}: The Eddington bias \citep{Eddington40} causes objects detected close to the sensitivity limit
  to be preferentially recorded with count-rates significantly above their true value; this can result in sources which are
  in reality below the historical upper limit (i.e.\ not transients) to be misclassified as transients. We are not able to say
  a-priori whether a given source is affected by this bias, although \cite{Evans14} (section 6.2 and fig.~10) showed
  that SXPS sources with fewer than $\til30$ counts are affected by this bias. Sources whose peak count-rate is
  $<3 \s$ above the $3-\s$ upper limit and which have fewer than 30 counts in the detection are usually classed
  as \emph{low significance}.

  \item {\bf Needs follow up}: This is somewhat of a catch-all category, where the source does not obviously fall into 
  either category above, but there is some doubt as to its nature and further observations are felt necessary before
  it can be definitively identified as a transient.

  \item {\bf Confirmed transient}: The object is definitely a new transient.
  
\end{itemize}

\noindent The reader will immediately realise that these classifications are to a considerable degree
subjective, and they should be interpreted as such. All events falling into these categories are
made publicly available (Section~\ref{sec:transAnnounce}) and the classifications are
intended as a guide and to enable filtering, rather than a definitive statement; this also helps
the clear candidates to be easily identifiable from among the (larger number of) low significance and
outburst events.

The \emph{outburst} category is particularly subjective, in that in reality the only phenomenological difference
between an outburst and a transient is the sensitivity of previous observations. Two identical sources, with quiescent
fluxes of (say) $10^{-14}$ erg s$^{-1}$ and outburst fluxes 100 times higher, could be classed one as a transient
and one as an outburst, if the latter had a historical observation long enough for forced photometry to detect it in
quiescence. 

\begin{figure}
  \begin{center}
  \includegraphics[width=8.1cm,angle=0]{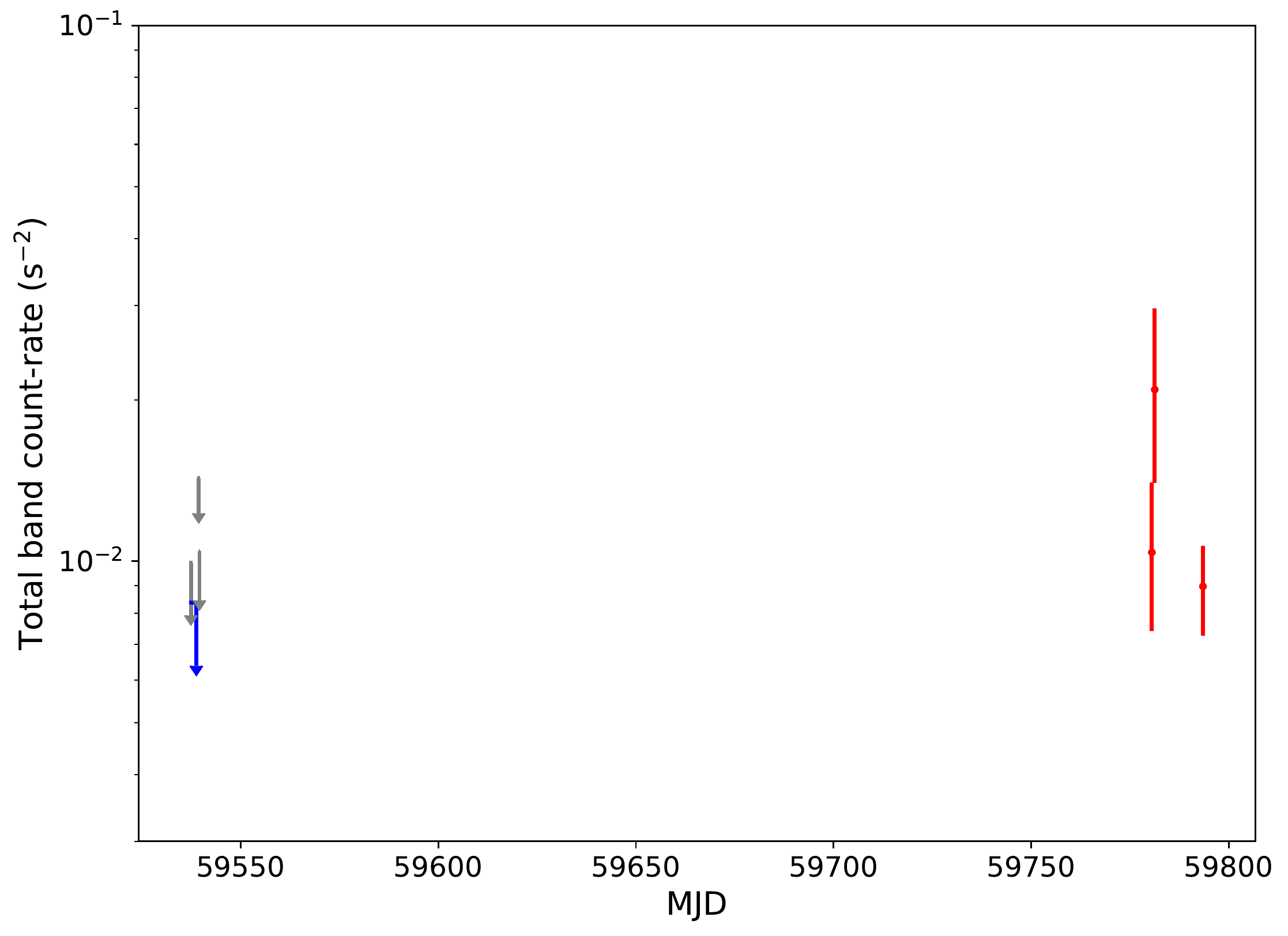}
  \end{center}
  \caption{0.3--10 keV light curve of Swift J051001.8-055931, one bin per \swift\ observation. The transient was
  discovered in the observation corresponding to the penultimate datapoint; the final datapoint comes from the follow-up
  ToO observations, i.e. this datapoint was not available when the transient was first reported to the XRT team.
  The grey upper limits are from the individual observations; the blue upper limit is the limit from the stacked image
  which combined these observations.}
\label{fig:demoTrans}
\end{figure}

Swift J051001.8-055931 presents a good case study of this subjectivity. It was discovered in
observation 00014892016 (version 4, LSXPS DatasetID: 225587) which began at 23:59 UT on 2022 July 20.
The light curve (Fig.~\ref{fig:demoTrans}) showed a detection in the previous observation which had
been accumulated over the previous 16 hours\footnote{Not accumulated continuously; that observation
contained 3.2 ks of exposure gathered over the 16-hour interval.}, and several non-detections (grey
arrows) more than 240 days earlier. At the time of the discovery, the LSXPS stacked image contained
only those non-detections, from which an upper limit of 0.0084 ct s$^{-1}$ was deduced (blue arrow);
the count-rate at discovery is clearly well above the upper limit, but the automated system
recommended this be classified as an outburst due to the detection in the previous observation. As
the day-old `historical' detection could indicate the rise of the transient, this was instead
manually classifiedas \emph{needs followup}. A \swift\ ToO observation was requested to investigate
further; this revealed that the source, while still detectable, had dropped below the historical
upper limit and thus the `transient' discovery was probably either a short-lived outburst or just a
low-significance fluctuation. In this case we classified it as an outburst, but \emph{low
significance} would be equally justifiable.

\subsection{Transient announcement and data access}
\label{sec:transAnnounce}

As soon as a transient is classified into one of the interesting classes (Section~\ref{sec:classify}),
it is published on the public transient webpage: \url{https://www.swift.ac.uk/LSXPS/transients}, and added
to the public transients database (Section~\ref{sec:contents}). At the present time, no push notifications are produced,
although we are discussing the recording of these transients (or some subset thereof) in the Transient Name Server.

If a transient is reobserved after discovery, whether intentionally or serendipitously, its light
curve is updated and also a second spectrum constructed and fitted, using all data from the
discovery observation onwards. Normally these products are only updated until the discovery dataset
enters the live LSXPS catalogue (Section~\ref{sec:build}), at which point the transient will
correspond to an LSXPS source (and the transient page and database will contain this link), and the
source products are kept up to date in the main catalogue (Section~\ref{sec:sourceProd}). However,
because source products are only updated to include observations that have been added to LSXPS, they
lag a few days behind real time. On occasion there may be a transient for which follow-up
observations have been requested and for which it is desireable to maintain up-to-date products in
real time. In such cases the XRT team will manually flag the transient and its products in the
transient area will continue to be updated each time new data are received. An example of such an
event is Swift J023017.0+283603, \rev{a source which demonstrates the power of our new transient detector.
It is consistent with the nucleus of a known galaxy and has a very soft X-ray spectrum, and so was
initially announced as a probable tidal disruption event \citep{ATEL15454}. Had this source only been found in later analysis,
as was hitherto typical, this classification would likely have been unchallenged. However, the prompt
transient alert from our new system allowed rapid and long-term multi-wavelength monitoring, exposing
variability on timescales of weeks, challenging the initial classification of the object and exposing
its true, and much more enigmatic, nature (Evans \etal2022, in prep).}

The transient data and database tables can be accessed through the LSXPS website and {\sc swifttools Python} module as described
in Section~\ref{sec:access}.

\section{Catalogue characteristics and contents}
\label{sec:contents}

\begin{figure}
  \begin{center}
  \includegraphics[width=8.1cm,angle=0]{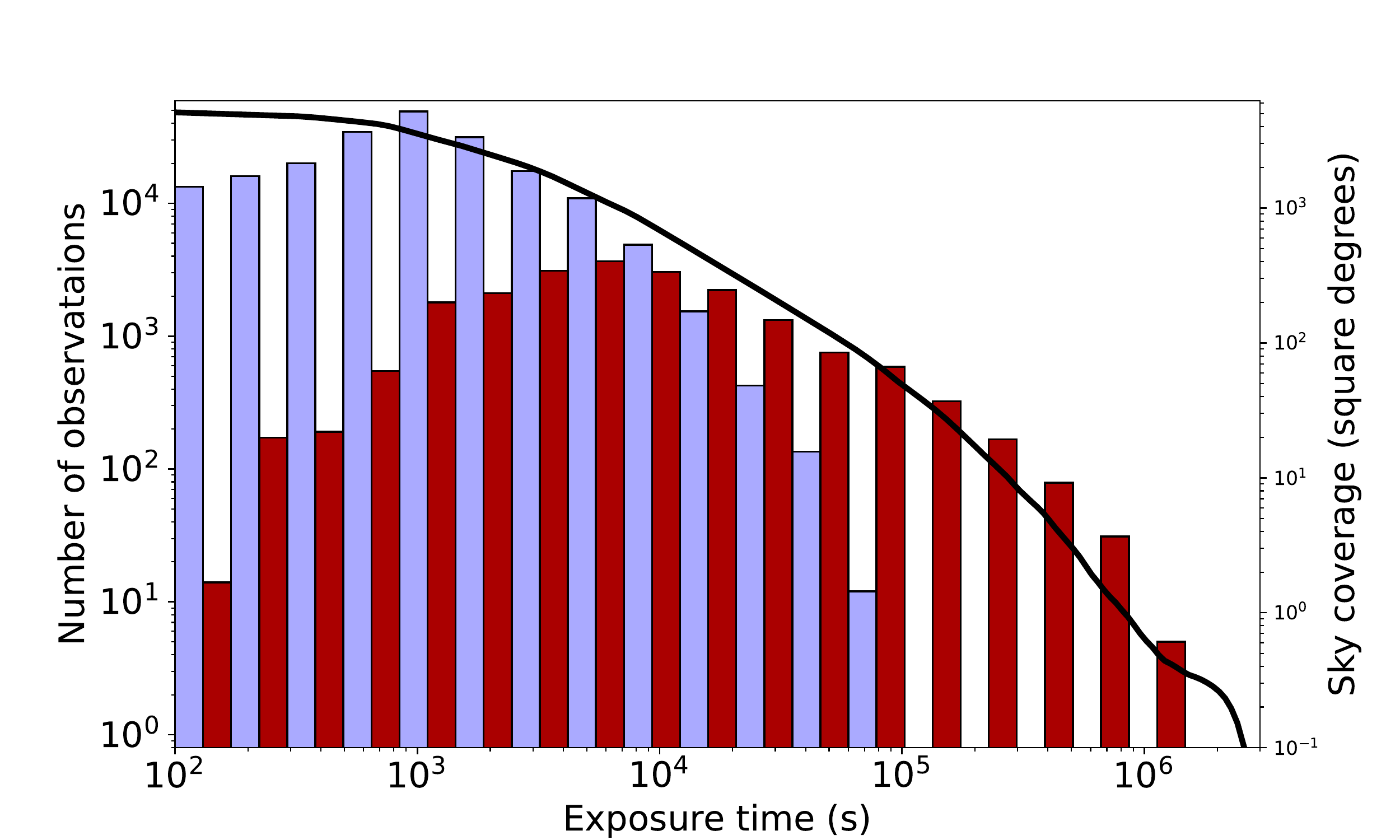}
  \end{center}
  \caption{The sky coverage and exposure details of the LSXPS catalog as of 2022 August 30. The solid line shows the sky coverage (corrected for overlaps)
  as a cumulative function of exposure time (i.e.\ area with at least the exposure indicated). The histogram shows the distribution of exposure time
  per dataset, with the individual observations shown in light blue and the stacked
  images in red; the different colors are each half the width of the actual bins.}
\label{fig:exposureArea}
\end{figure}

Being a dynamic catalogue, any statement of characteristics is obviously tied to the time at which they were generated: the `live' 
status of the numbers given here are available on the LSXPS website. The numbers given below were
obtained at 15:00 UT on 2022 August 30.

The LSXPS catalogue contains 279,021 unique sources, composed from 1.5 million blind detections across all
four energy bands, or 756,769 `obs sources'\footnote{An `obs source' is produced by merging all
detections of the same object across the energy bands within a single dataset; this is analogous to
what the \xmm\ catalogues call a detection.}. The median 0.3--10 keV flux (assuming a power-law
spectrum) is $4.1\tim{-14}$ \ergcms. The sky coverage of LSXPS is shown in
Fig.~\ref{fig:exposureArea}; the total unique sky coverage is 5,186 deg$^2$, 3,503 deg$^2$ have at
least 1 ks of exposure and $\til$700 deg$^2$ have been observed with at least 10 ks. The 2SXPS
catalogue contained data up to 2018 August 1; thus on average LSXPS has grown by 49 sources per day
and the unique sky coverage is increasing at 0.94 deg$^2$/day.

The source detection mechanism employed for LSXPS was the same as for 2SXPS, and so only summarised
here. For full details see \cite{Evans20}, especially sections 3.5 and 7, figs.~6--7 and tables
~3~and~6.

As for 2SXPS, we provide `clean' and `ultra-clean' subsets of the sources and dataset. Clean sources
are those with a best detection flag of 0 or 1 (i.e. \emph{Good} or \emph{Reasonable} with no other
warning bits set); OpticalLoadingWarning, StrayLightWarning and NearBrightSourceWarning all unset;
and a field flag of 0 or 1 (see Table~\ref{tab:fieldflag}). Ultra-clean sources are  a subset of the
clean sources, with detection and field flags of 0. There are 190,902 clean sources and 170,372
ultra-clean sources in LSXPS, a growth rate of 30 (26) per day for the clean (ultra-clean) classes.

The actual catalogue contents are organised into six publicly-available tables, which are described in 
Appendix~B.

\section{Access to LSXPS: the website and API}
\label{sec:access}

The LSXPS catalogue is available from the UKSSDC website via: {\url{https://www.swift.ac.uk/LSXPS}.
This website provides interactive tools to explore the catalogue (i.e. cone searches, filters, etc.)
and visualisations of all datasets and sources and relevant properties. The catalogue tables
(Appendix~B) can also be downloaded from here as FITS or CSV files. Due to the
dynamic nature of the catalogue, these files are updated every hour. These hourly snapshots are
available via the website for one week after creation;  daily snapshots are created at midnight UT
each day and available indefinitely. It is thus recommended that for any serious project based on a
frozen snapshot of LSXPS, one of these daily snapshots is used to enable the work to be reproducible
in future.

The website also includes an upper limit calculator, as provided for the previous catalogues; the
operation of this system was described in \cite{Evans14}, section~4.4. For LSXPS this server queries
all data, including those still in the cache (that is, observations which are not marked as
complete, see Section~\ref{sec:LSXPSdata}). The ESA HILIGT upper limit server
\citep{Saxton22,Konig22} currently queries 2SXPS for upper limits from \swift, but this will soon be
updated to use LSXPS.

The catalogue can also be accessed through the {\sc swifttools Python} module. This module, which
can be installed via {\sc pip}, contains two components, both of which are extensively documented.
{\sc swifttools.swift\_too} is maintained by Jamie Kennea (at Penn State Univerity) and provides functionality including access
to \swift\ data, the ability to query the observability and observing history of a given object, and to
submit ToO requests; it is documented at \url{https://www.swift.psu.edu/too\_api/}. {\sc
swifttools.ukssdc} is maintained by Phil Evans, and was added in {\sc swifttools} v3.0. This version
includes tools to access data and query different \swift\ catalogues; this includes GRB data and catalogues,
and LSXPS. All functionality of the LSXPS website, including the upper limit server is available
through the API, including integration with the tools that allow on-demand analysis of sources
(themselves provided through the {\sc swifttools.ukssdc.xrt\_prods} module\footnote{Prior to {\sc
swifttools} v3.0, this module was {\sc swifttools.xrt\_prods}, and while that path is still active
and supported, it now simply silently wraps {\sc swifttools.ukssdc.xrt\_prods}. There are excellent
reasons for this, but they are very boring.}), and much of the functionality has also been backported
to 2SXPS, so that this can also be queried by the module. This module is documented at \url{https://www.swift.ac.uk/API/}.

We ask that users of the catalogue, via the website, data tables or API, acknowledge this in their
subsequent publications: by citing this paper and also including in the acknowledgements the sentence:
\emph{This work made use of data supplied by the UK Swift Science Data Centre at the University of Leicester.}

\section{Summary}

\rev{We have presented the first-ever low-latency, sensitive X-ray point-source catalogue and transient
detector. This enables rapid follow up of X-ray transients too faint to trigger all-sky monitors and
so opens up a new region of transient phase-space for exploration. This is enabled by the unique
provision of a `living' source catalogue, LSXPS; which is kept constantly up to date every time a new
\swift\ observation is marked as completed. This catalogue and associated upper limit server
provide not only a reference for our transient detector, but a valuable resource for the rapidly-growing
field of time-domain and multi-messenger astronomy generally, in which it is often useful to have quick access
to the most up-to-date analysis of a given sky location.}






\section*{Acknowledgements}
For the purpose of open access, the author has applied a Creative Commons Attribution (CC BY)
licence to any Author Accepted Manuscript version arising. This work made use of data supplied by
the UK Swift Science Data Centre at the University of Leicester. PAE, KLP, ABP and RAJE-F
acknowledge UKSA support. SC acknowledges the support from the Italian Space Agency, contract
ASI/INAF n. I/004/11/5. This work is partially supported by a grant from the Italian Ministry of
Foreign Affairs and International Cooperation Nr. MAE0065741.

\section*{Data availability}

All of the original \swift\ data are available via the three Swift datacentres:
(\url{https://www.swift.ac.uk/swift_live/}, \url{https://swift.gsfc.nasa.gov/archive},
\url{https://www.ssdc.asi.it/mmia/index.php?mission=swiftmastr}). The data derived for this paper
are available through the LSXPS catalogue web pages {\url{https://www.swift.ac.uk/LSXPS/}) and can
also be accessed using the {\sc swifttools} Python module, available via {\sc pip}.

\bibliographystyle{mnras} \bibliography{phil}

\onecolumn
\appendix
\section{Dataset processing algorithm}
\label{sec:appBuild}

This appendix contains an overview  of the processing of datasets in LSXPS. For a full description
including details of how the various tasks are performed, see \cite{Evans20}: particularly
sections~2--3 and (for details of stray light modelling) appendix~A. 

The first step of dataset processing differs for single observations and stacked images. For a
single observation, the data are filtered to remove all times potentially contaminated by bright
Earth or when the spacecraft astrometry was unstable or unreliable. If this leaves less than 100-s
of cleaned PC-mode data, the dataset is discarded and the processing stops -- the dataset is not
added to LSXPS. Otherwise, the data are split up into snapshots -- times of continuous observing. This is
necessary because \swift\ does not point in exactly the same direction each time it observes a given
target. Snapshots with less than 50~s of PC-mode exposure are rejected (and if no snapshot has
sufficient exposures, the observation is discarded); for the other snapshots the pointing direction
and XRT window size are recorded, and a catalogue check is carried out to identify any bright X-ray
sources outside the field of view that could potentially cause stray light patterns. Images are
created for each snapshot in each of the four LSXPS energy bands, and an exposure map and total-band
event list are also created. These per-snapshot results are then combined to create a summed image
per energy band and a summed exposure map and event lists (the individual images and exposure maps
are retained, as these are needed by the source detection code).

\begin{figure*}
  \begin{center}
  \includegraphics[width=16cm,angle=0]{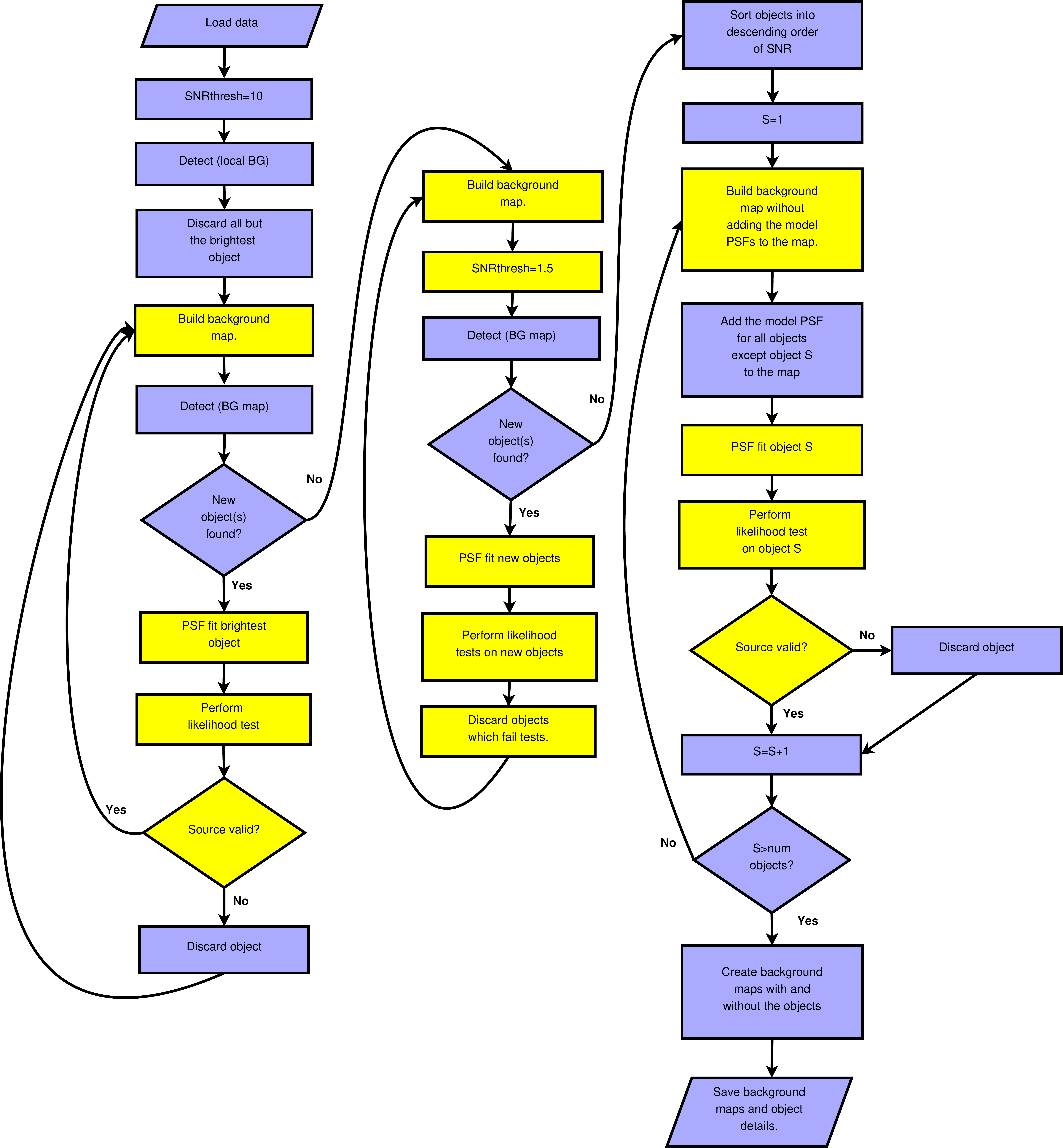}
  \end{center}
  \caption{Diagrammatic outline of the source detection mechanism, reproduced from fig~2 of Evans \etal(2020)}.
\label{fig:algorithm}
\end{figure*}

Stacked images are built from individual observations in LSXPS, thus the steps above do not need to
be repeated. The per-snapshot images, event lists and exposure maps are simply retrieved for each
observation in the stack, these are then mapped onto the WCS coordinate system of the stacked image
and summed. From this point onwards the processing is identical for observations and stacked images. 

The next step is source detection and characterisation: the algorithm for this is shown in
Fig.~\ref{fig:algorithm}. For each energy band, possible sources are identified by a sliding cell
approach, and then localised with a PSF fit; likelihood tests are carried out to either rank them as
\emph{Good}, \emph{Reasonable} or \emph{Poor}, or to reject them as spurious. This process is
iterative, with a background model constructed in each cycle using the information from the previous
cycles; this includes adding a PSF model of each source identified so far, reducing the number of
aliases and helping to deconvolve nearby sources. This background map also includes modelling any
stray light in the image\footnote{Stray light is only freely fitted to the total-band image of
single observations. For the sub-bands the position of the stray-light source is fixed, but the
normalisation can vary. For stacked images the stray light definitions are taken from the individual
observations' results, and only the normalisation is free to vary, even in the total band}. Once the
final list of objects in the image has been produced, their positions are all redetermined (in
descending order of S/N ratio) since these may be improved with full knowledge of the other sources
in the image. The count-rates for each source in this image (and hence energy band) is also
calculated, unlike in many X-ray catalogues, these are fully corrected for effects such as pile-up,
vignetting, dead columns or pixels on the CCD etc.

There are two minor changes to the details of this process since 2SXPS. The first of these is a
recalibration of the point-spread-function (PSF). This had been done for 2SXPS and the new PSF
was described in appendix B of \cite{Evans20}; however some small improvements were made some time
after the publication of that work, and a new PSF calibration file was formally released into the
CALDB by the \swift\
team\footnote{\url{https://heasarc.gsfc.nasa.gov/docs/heasarc/caldb/swift/docs/xrt/SWIFT-XRT-CALDB-10\_v01.pdf}}.
The changes were negligible; however, we adopted the officially-released PSF for this catalogue.

The second change relates to the way in which pile up is handled in the partial energy bands. When a
source is piled up the PSF will be distorted in all energy bands, regardless of the count-rate in
that band: an absorbed source can have a very piled-up PSF and yet a very low count-rate in the soft
band, for example. To ensure that the PSF was properly fitted in 2SXPS, we required that sources in
the sub-bands that were likely counterparts to a piled-up source in the total band, be fitted with
the piled up PSF (section 3.4 of \citealt{Evans20}). This sometimes still gave poor PSF
characterisation or position measurement in the sub-bands, so for LSXPS we have modified this
further, constraining the pile-up parameters of the PSF model in the sub-bands to be fixed at the
values determined in the total band. It is still the case, as in 2SXPS, that if a piled up source is not found at all
in a sub-band, but other, non-piled-up sources are found near to the positon of that source in the total band,
those sources are flagged as likely aliases of the piled up source, and bit 3 (value 8) of their detection flag
is set (Table~\ref{tab:detwarn}).

Once source detection is complete, the source lists from the different energy bands are combined
to make a unique list of objects found in this image; this is based on spatial separation and does not
use the systematic error on the XRT astrometry, since it is common to all bands. Once the list of unique
objects in the dataset has been created, checks are carried out to warn if any of them are likely to be
affected by optical loading, stray light or are likely aliases of a bright source.

In 2SXPS, this marked the end of the individual dataset processing. For LSXPS, the dataset is
then subjected to the transient search process detailed in Section~\ref{sec:transCheck}, and then the results
are stored in the LSXPS cache.

\section{Catalogue tables}
\label{sec:appCats}

There are seven tables available for download, the contents of which are described below.
These tables are:

  

\begin{itemize}
  \item Sources
  \item Datasets
  \item Detections
  \item ObsSources
  \item External catalogue matches
  \item Old Stacks
  \item Transients
\end{itemize}

In the following sections we give for each a short description and then a list of all of the table columns.

Many of the properties in the tables have errors associated with them; these are given in two extra
columns with `\_pos' and `\_neg' added to the column names, e.g. the column `Rate' is followed by
`Rate\_pos' and `Rate\_neg'. Rather than list all of these columns (the tables are long
enough\ldots), the tables below include the field `Has Errors?' If this is `Yes' it indicates that
these extra columns exist as well. These errors are all 1-\s\ errors and assume a Gaussian
distribution. A value of `[bool]' in the units column indicates that the field is boolean.

\subsection{Sources}

The `Sources' table is the main LSXPS table, containing details of all of the unique sources in LSXPS. Note that sources can be
removed from this table as the catalogue evolves, as well as being added to it (see Section~\ref{sec:addSources}).


\begin{longtable}{cclc}
  \caption{Contents of the `Sources' database table.}
  \label{tab:sources} \\
  \hline
Field & Units & Description & Has Errors? \\
  \hline
  \vspace{0.1cm}
  \endfirsthead
  \multicolumn{4}{c}%
{{\bfseries \tablename\ \thetable{} -- continued from previous page}} \\
\hline
Field & Units & Description & Has Errors? \\
\hline
\vspace{0.1cm}
\endhead
\multicolumn{4}{r}{{\emph{Continued\ldots}}} \\ \hline
\endfoot
\hline
\endlastfoot

LSXPS\_ID                &                                & Numerical unique source identifier within LSXPS.   &  \\
IAUName                  &                                & IAU-format name, LSXPS JHHMMSS+-ddmmsss.           &  \\
\multicolumn{4}{c}{\emph{Name and position}} \\
RA                       & Deg                            & Right Ascension (J2000) in decminal degrees.       &  \\
Decl                     & Deg                            & Declination (J2000) in decimal degrees.            &  \\
Err90                    & arcsec                         & Position uncertainty, 90\% confidence, radial,     &  \\
                         &                                & assumed to be Rayleigh-distributed.                &  \\
AstromType               &                                & Provenance of source astrometry. 0=Swift star      &  \\
                         &                                & tracker, 1=XRT/2MASS astrometry.                   &  \\
l                        & Deg                            & Galactic longitude.                                &  \\
b                        & Deg                            & Galactic latitude.                                 &  \\
MeanOffAxisAngle         & arcmin                         & The mean angular distance of the source from the   &  \\
                         &                                & XRT boresight in all observations within which     &  \\
                         &                                & the source was detected (arcmin).                  &  \\
NearestNeighbour         & arcsec                         & The distance  to the closest LSXPS source to this  &  \\
                         &                                & one.                                               &  \\
NearestOKNeighbour       & arcsec                         & The distance   to the closest LSXPS source to      &  \\
                         &                                & this one which is ranked Good or Reasonable and    &  \\
                         &                                & has no other DetFlag bits set.                     &  \\
NearestNeighbour\_ID     &                                & The LSXPS\_ID of the closest LSXPS source to this  &  \\
                         &                                & one.                                               &  \\
NearestOKNeighbour\_ID   &                                & The LSXPS\_ID of the closest LSXPS source to this  &  \\
                         &                                & one which is ranked Good or Reasonable and has no  &  \\
                         &                                & other DetFlag bits set.                            &  \\
HPPIX                    &                                & The Healpix pixel containing the source            &  \\
                         &                                & (NSIDE=128, ordering=RING).                        &  \\
\multicolumn{4}{c}{\emph{Exposure details}} \\
Exposure                 & s                              & The total exposure at the source location in the   &  \\
                         &                                & catalogue.                                         &  \\
FirstObsDate             &                                & The UTC date \& time of the start of the first     &  \\
                         &                                & observation in LSXPS which covered the source      &  \\
                         &                                & location.                                          &  \\
LastObsDate              &                                & The UTC date \& time of the end of the last        &  \\
                         &                                & observation in LSXPS which covered the source      &  \\
                         &                                & location.                                          &  \\
FirstObsMET              & MET                            & The time of the start of the first observation in  &  \\
                         &                                & LSXPS which covered the source location, in Swift  &  \\
                         &                                & Mission Elapsed Time.                              &  \\
LastObsMET               & MET                            & The time of the end of the last observation in     &  \\
                         &                                & LSXPS which covered the source location in, Swift  &  \\
                         &                                & Mission Elapsed Time.                              &  \\
FirstDetDate             &                                & The UTC date \& time of the start of the first     &  \\
                         &                                & observation in LSXPS in which the source count     &  \\
                         &                                & rate is inconsisent with 0 at the 3-\s level.      &  \\
LastDetDate              &                                & The UTC date \& time of the end of the last        &  \\
                         &                                & observation in LSXPS in which the source count     &  \\
                         &                                & rate is inconsisent with 0 at the 3-\s level.      &  \\
FirstDetMET              & MET                            & The time of the start of the first observation in  &  \\
                         &                                & LSXPS in which the source count rate is            &  \\
                         &                                & inconsisent with 0 at the 3-\s level, in Swift     &  \\
                         &                                & Mission Elapsed Time.                              &  \\
LastDetMET               & MET                            & The time of the end of the last observation in     &  \\
                         &                                & LSXPS in which the source count rate is            &  \\
                         &                                & inconsisent with 0 at the 3-\s level, in Swift     &  \\
                         &                                & Mission Elapsed Time.                              &  \\
FirstBlindDetDate        &                                & The UTC date \& time of the start of the first     &  \\
                         &                                & observation in LSXPS in which the source is        &  \\
                         &                                & detected in the blind search.                      &  \\
LastBlindDetDate         &                                & The UTC date \& time of the end of the last        &  \\
                         &                                & observation in LSXPS in which the source is        &  \\
                         &                                & detected in the blind search.                      &  \\
FirstBlindDetMET         & MET                            & The time of the start of the first observation in  &  \\
                         &                                & LSXPS in which the source is detected in the       &  \\
                         &                                & blind search, in Swift Mission Elapsed Time.       &  \\
LastBlindDetMET          & MET                            & The time of the end of the last observation in     &  \\
                         &                                & LSXPS in which the source is detected in the       &  \\
                         &                                & blind search, in Swift Mission Elapsed Time.       &  \\
NumObs                   &                                & The number of observations covering the position   &  \\
                         &                                & of this source.                                    &  \\
NumBlindDetObs           &                                & The number of observations in which this source    &  \\
                         &                                & was found in a blind search.                       &  \\
NumDetObs                &                                & The number of observations in which this source    &  \\
                         &                                & is detected.                                       &  \\
BestDetectionID          &                                & The ID of the 'best' detection, (cf the            &  \\
                         &                                & detections table).                                 &  \\
NonBlindDet\_band0       &                                & Whether the count-rate in the 0.3--10 keV band is  &  \\
                         &                                & inconsistent with 0 at the 3-\s level.             &  \\
NonBlindDet\_band1       &                                & Whether the count-rate in the 0.3--1 keV band is   &  \\
                         &                                & inconsistent with 0 at the 3-\s level.             &  \\
NonBlindDet\_band2       &                                & Whether the count-rate in the 1-2 keV band is      &  \\
                         &                                & inconsistent with 0 at the 3-\s level.             &  \\
NonBlindDet\_band3       &                                & Whether the count-rate in the 2--10 keV band is    &  \\
                         &                                & inconsistent with 0 at the 3-\s level.	            &  \\
\multicolumn{4}{c}{\emph{Quality flag parameters}} \\
DetFlag                  &                                & The final source detection flag.                   &  \\
FieldFlag                &                                & The final fieldFlag for this source.               &  \\
DetFlag\_band0           &                                & The final detection flag the 0.3--10 keV band.     &  \\
DetFlag\_band1           &                                & The final detection flag the 0.3--1 keV band.      &  \\
DetFlag\_band2           &                                & The final detection flag in the 1-2 keV band.      &  \\
DetFlag\_band3           &                                & The final detection flag in the 2--10 keV band.    &  \\
OpticalLoadingWarning    &                                & The worst optical loading warning from all         &  \\
                         &                                & detections of this source. If this is 0 there is   &  \\
                         &                                & no warning, otherwise there is a nearby optical    &  \\
                         &                                & source brighter than the level at which optical    &  \\
                         &                                & loading becomes a concern; the value indicates     &  \\
                         &                                & how many magnitudes brighter than that level the   &  \\
                         &                                & source is.                                         &  \\
StrayLightWarning        &                                & Whether any detection of this source occurred      &  \\
                         &                                & within fitted stray light rings.                   &  \\
NearBrightSourceWarning  &                                & Whether any detection of this source occurred      &  \\
                         &                                & within the PSF wings of a bright object.           &  \\
IsPotentialAlias         &                                & Whether this source is potentially an alias (i.e.  &  \\
                         &                                & duplicate) of another source.                      &  \\
\multicolumn{4}{c}{\emph{Processing details}} \\
WhenAdded                &                                & The date at which this source was added to LSXPS.  &  \\
StillDetected            &                                & Whether the source was detected in the most        &  \\
                         &                                & recent observation of its location (1=yes; 0=no,   &  \\
                         &                                & but last upper limit unconstraining; -1 = no, and  &  \\
                         &                                & upper limit below last detection).                 &  \\
ProcessedStatus          &                                & A bit-wise flag describing the processing status   &  \\
                         &                                & of the source data.                                &  \\
WhenModified             &                                & The date at which this source was last analysed.   &  \\
\multicolumn{4}{c}{\emph{Count rate and variability measurements}} \\
Rate\_band0              & s$^{-1}$                       & The mean count-rate in the 0.3--10 keV band.       & Yes \\
HR1                      &                                & The aggregate HR1 hardness ratio of the source.    & Yes \\
HR2                      &                                & The aggregate HR2 hardness ratio of the source.    & Yes \\
Rate\_band1              & s$^{-1}$                       & The mean count-rate in the 0.3--1 keV band.        & Yes \\
Rate\_band2              & s$^{-1}$                       & The mean count-rate in the 1-2 keV band.           & Yes \\
Rate\_band3              & s$^{-1}$                       & The mean count-rate in the 2--10 keV band.         & Yes \\
Counts\_band0            &                                & The total number of counts in the source region    &  \\
                         &                                & in the 0.3--10 keV band.                           &  \\
Counts\_band1            &                                & The total number of counts in the source region    &  \\
                         &                                & in the 0.3--1 keV band.                            &  \\
Counts\_band2            &                                & The total number of counts in the source region    &  \\
                         &                                & in the 1-2 keV band.                               &  \\
Counts\_band3            &                                & The total number of counts in the source region    &  \\
                         &                                & in the 2--10 keV band.                             &  \\
BgCounts\_band0          &                                & The total number of background counts expected in  &  \\
                         &                                & the source region in the 0.3--10 keV band.         &  \\
BgCounts\_band1          &                                & The total number of background counts expected in  &  \\
                         &                                & the source region in the 0.3--1 keV band.          &  \\
BgCounts\_band2          &                                & The total number of background counts expected in  &  \\
                         &                                & the source region in the 1-2 keV band.             &  \\
BgCounts\_band3          &                                & The total number of background counts expected in  &  \\
                         &                                & the source region in the 2--10 keV band.           &  \\
RateCF\_band0            & s$^{-1}$                       & The PSF correction factor in the 0.3--10 keV       &  \\
                         &                                & band.                                              &  \\
RateCF\_band1            & s$^{-1}$                       & The PSF correction factor in the 0.3--1 keV band.  &  \\
RateCF\_band2            & s$^{-1}$                       & The PSF correction factor in the 1-2 keV band.     &  \\
RateCF\_band3            & s$^{-1}$                       & The PSF correction factor in the 2--10 keV band.   &  \\
UL\_band0                & s$^{-1}$                       & The 3-\s upper confidence limit on the count-rate  &  \\
                         &                                & in the 0.3--10 keV band.                           &  \\
UL\_band1                & s$^{-1}$                       & The 3-\s upper confidence limit on the count-rate  &  \\
                         &                                & in the 0.3--1 keV band.                            &  \\
UL\_band2                & s$^{-1}$                       & The 3-\s upper confidence limit on the count-rate  &  \\
                         &                                & in the 1-2 keV band.                               &  \\
UL\_band3                & s$^{-1}$                       & The 3-\s upper confidence limit on the count-rate  &  \\
                         &                                & in the 2--10 keV band.                             &  \\
PeakRate\_band0          & s$^{-1}$                       & The peak count-rate in the 0.3--10 keV band.       & Yes \\
PeakRate\_band1          & s$^{-1}$                       & The peak count-rate in the 0.3--1 keV band.        & Yes \\
PeakRate\_band2          & s$^{-1}$                       & The peak count-rate in the 1-2 keV band.           & Yes \\
PeakRate\_band3          & s$^{-1}$                       & The peak count-rate in the 2--10 keV band.         & Yes \\
\multicolumn{4}{c}{\emph{Flux measurements and spectral parameters}} \\
GalacticNH               & \cms                           & The Galactic absorption column in the direction    &  \\
                         &                                & of the source, from Willingale et al (2013), in    &  \\
                         &                                & \cms.                                              &  \\
PvarPchiSnapshot\_band0  &                                & The probability that the source count-rate in the  &  \\
                         &                                & 0.3--10 keV band does not vary between snapshots.  &  \\
PvarPchiSnapshot\_band1  &                                & The probability that the source count-rate in the  &  \\
                         &                                & 0.3--1 keV band does not vary between snapshots.   &  \\
PvarPchiSnapshot\_band2  &                                & The probability that the source count-rate in the  &  \\
                         &                                & 1-2 keV band does not vary between snapshots.      &  \\
PvarPchiSnapshot\_band3  &                                & The probability that the source count-rate in the  &  \\
                         &                                & 2--10 keV band does not vary between snapshots.    &  \\
PvarPchiSnapshot\_HR1    &                                & The probability that the source HR1 hardness       &  \\
                         &                                & ratio does not vary between snapshots.             &  \\
PvarPchiSnapshot\_HR2    &                                & The probability that the source HR2 hardness       &  \\
                         &                                & ratio does not vary between snapshots.             &  \\
PvarPchiObsID\_band0     &                                & The probability that the source count-rate in the  &  \\
                         &                                & 0.3--10 keV band does not vary between             &  \\
                         &                                & observations.                                      &  \\
PvarPchiObsID\_band1     &                                & The probability that the source count-rate in the  &  \\
                         &                                & 0.3--1 keV band does not vary between              &  \\
                         &                                & observations.                                      &  \\
PvarPchiObsID\_band2     &                                & The probability that the source count-rate in the  &  \\
                         &                                & 1-2 keV band does not vary between observations.   &  \\
PvarPchiObsID\_band3     &                                & The probability that the source count-rate in the  &  \\
                         &                                & 2--10 keV band does not vary between               &  \\
                         &                                & observations.                                      &  \\
PvarPchiObsID\_HR1       &                                & The probability that the source HR1 hardness       &  \\
                         &                                & ratio does not vary between observations.          &  \\
PvarPchiObsID\_HR2       &                                & The probability that the source HR2 hardness       &  \\
                         &                                & ratio does not vary between observations.          &  \\
\multicolumn{4}{c}{\emph{Flux measurements and spectral parameters}} \\
WhichPow                 &                                & Which method of determining the spectral           &  \\
                         &                                & properties assuming a power-law was used:          &  \\
                         &                                & 0=Canned spectrum, 1=Interpolated from the HR,     &  \\
                         &                                & 2=Spectrum constructed and fitted.                 &  \\
WhichAPEC                &                                & Which method of determining the spectral           &  \\
                         &                                & properties assuming an APEC was used: 0=Canned     &  \\
                         &                                & spectrum, 1=Interpolated from the HR, 2=Spectrum   &  \\
                         &                                & constructed and fitted.                            &  \\
PowECFO                  & \ecf                           & The Energy Conversion Factor from 0.3--10 keV      &  \\
                         &                                & count-rate to 0.3--10 keV observed flux, assuming  &  \\
                         &                                & a power-law spectrum.                              &  \\
PowECFU                  & \ecf                           & The Energy Conversion Factor from 0.3--10 keV      &  \\
                         &                                & count-rate to 0.3--10 keV unabsorbed flux,         &  \\
                         &                                & assuming a power-law spectrum.                     &  \\
PowFlux                  & \ergcms                        & The mean 0.3--10 keV observed flux assuming a      & Yes \\
                         &                                & power-law spectrum.                                & Yes \\
PowUnabsFlux             & \ergcms                        & The mean 0.3--10 keV unabsorbed flux assuming a    & Yes \\
                         &                                & power-law spectrum.                                & Yes \\
APECECFO                 & \ecf                           & The Energy Conversion Factor from 0.3--10 keV      &  \\
                         &                                & count-rate to 0.3--10 keV observed flux, assuming  &  \\
                         &                                & an APEC spectrum.                                  &  \\
APECECFU                 & \ecf                           & The Energy Conversion Factor from 0.3--10 keV      &  \\
                         &                                & count-rate to 0.3--10 keV unabsorbed flux,         &  \\
                         &                                & assuming an APEC spectrum.                         &  \\
APECFlux                 & \ergcms                        & The mean 0.3--10 keV observed flux assuming an     & Yes \\
                         &                                & APEC spectrum.                                     & Yes \\
APECUnabsFlux            & \ergcms                        & The mean 0.3--10 keV unabsorbed flux assuming an   & Yes \\
                         &                                & APEC spectrum.                                     & Yes \\
PowPeakFlux              & \ergcms                        & The peak 0.3--10 keV observed flux assuming a      & Yes \\
                         &                                & power-law spectrum.                                & Yes \\
PowPeakUnabsFlux         & \ergcms                        & The peak 0.3--10 keV unabsorbed flux assuming a    & Yes \\
                         &                                & power-law spectrum.                                & Yes \\
APECPeakFlux             & \ergcms                        & The peak 0.3--10 keV observed flux assuming an     & Yes \\
                         &                                & APEC spectrum.                                     & Yes \\
APECPeakUnabsFlux        & \ergcms                        & The peak 0.3--10 keV unabsorbed flux assuming an   & Yes \\
                         &                                & APEC spectrum.                                     & Yes \\
FixedPowECFO             & \ecf                           & The Energy Conversion Factor from 0.3--10 keV      &  \\
                         &                                & count-rate to 0.3--10 keV observed flux, assuming  &  \\
                         &                                & the canned power-law spectrum.                     &  \\
FixedPowECFU             & \ecf                           & The Energy Conversion Factor from 0.3--10 keV      &  \\
                         &                                & count-rate to 0.3--10 keV unabsorbed flux,         &  \\
                         &                                & assuming the canned power-law spectrum.            &  \\
FixedPowFlux             & \ergcms                        & The mean 0.3--10 keV observed flux assuming the    & Yes \\
                         &                                & canned power-law spectrum.                         & Yes \\
FixedPowUnabsFlux        & \ergcms                        & The mean 0.3--10 keV unabsorbed flux assuming the  & Yes \\
                         &                                & canned power-law spectrum.                         & Yes \\
FixedAPECECFO            & \ecf                           & The Energy Conversion Factor from 0.3--10 keV      &  \\
                         &                                & count-rate to 0.3--10 keV observed flux, assuming  &  \\
                         &                                & the canned APEC spectrum.                          &  \\
FixedAPECECFU            & \ecf                           & The Energy Conversion Factor from 0.3--10 keV      &  \\
                         &                                & count-rate to 0.3--10 keV unabsorbed flux,         &  \\
                         &                                & assuming the canned APEC spectrum.                 &  \\
FixedAPECFlux            & \ergcms                        & The mean 0.3--10 keV observed flux assuming the    & Yes \\
                         &                                & canned APEC spectrum.                              & Yes \\
FixedAPECUnabsFlux       & \ergcms                        & The mean 0.3--10 keV unabsorbed flux assuming the  & Yes \\
                         &                                & canned APEC spectrum.                              & Yes \\
InterpPowECFO            & \ecf                           & The Energy Conversion Factor from 0.3--10 keV      &  \\
                         &                                & count-rate to 0.3--10 keV observed flux, assuming  &  \\
                         &                                & the power-law spectrum interpolated from the HRs.  &  \\
InterpPowECFU            & \ecf                           & The Energy Conversion Factor from 0.3--10 keV      &  \\
                         &                                & count-rate to 0.3--10 keV unabsorbed flux,         &  \\
                         &                                & assuming the power-law spectrum interpolated from  &  \\
                         &                                & the HRs.                                           &  \\
InterpPowNH              & \cms                           & The hydrogen column density inferred assuming the  & Yes \\
                         &                                & power-law spectrum interpolated from the HRs.      & Yes \\
InterpPowGamma           &                                & The spectral photon index inferred assuming the    & Yes \\
                         &                                & power-law spectrum interpolated from the HRs.      & Yes \\
InterpPowFlux            & \ergcms                        & The mean 0.3--10 keV observed flux assuming the    & Yes \\
                         &                                & power-law spectrum interpolated from the HRs.      & Yes \\
InterpPowUnabsFlux       & \ergcms                        & The mean 0.3--10 keV unabsorbed flux assuming the  & Yes \\
                         &                                & power-law spectrum interpolated from the HRs.      & Yes \\
InterpAPECECFO           & \ecf                           & The Energy Conversion Factor from 0.3--10 keV      &  \\
                         &                                & count-rate to 0.3--10 keV observed flux, assuming  &  \\
                         &                                & the APEC spectrum interpolated from the HRs.       &  \\
InterpAPECECFU           & \ecf                           & The Energy Conversion Factor from 0.3--10 keV      &  \\
                         &                                & count-rate to 0.3--10 keV unabsorbed flux,         &  \\
                         &                                & assuming the APEC spectrum interpolated from the   &  \\
                         &                                & HRs.                                               &  \\
InterpAPECNH             & \cms                           & The hydrogen column density inferred assuming the  & Yes \\
                         &                                & APEC spectrum interpolated from the HRs.           & Yes \\
InterpAPECkT             & keV                            & The temperature inferred assuming the APEC         & Yes \\
                         &                                & spectrum interpolated from the HRs.                & Yes \\
InterpAPECFlux           & \ergcms                        & The mean 0.3--10 keV observed flux assuming the    & Yes \\
                         &                                & APEC spectrum interpolated from the HRs.           & Yes \\
InterpAPECUnabsFlux      & \ergcms                        & The mean 0.3--10 keV unabsorbed flux assuming the  & Yes \\
                         &                                & APEC spectrum interpolated from the HRs.           & Yes \\
P\_pow                   &                                &  The probability that the HR values of this        &  \\
                         &                                & source could be obtained if the true spectrum is   &  \\
                         &                                & an absorbed power-law.                             &  \\
P\_APEC                  &                                &  The probability that the HR values of this        &  \\
                         &                                & source could be obtained if the true spectrum is   &  \\
                         &                                & an APEC.                                           &  \\
FittedPowECFO            & \ecf                           & The Energy Conversion Factor from 0.3--10 keV      &  \\
                         &                                & count-rate to 0.3--10 keV observed flux, assuming  &  \\
                         &                                & the power-law spectral model fitted to a           &  \\
                         &                                & custom-built spectrum.                             &  \\
FittedPowECFU            & \ecf                           & The Energy Conversion Factor from 0.3--10 keV      &  \\
                         &                                & count-rate to 0.3--10 keV unabsorbed flux,         &  \\
                         &                                & assuming the power-law spectral model fitted to a  &  \\
                         &                                & custom-built spectrum.                             &  \\
FittedPowNH              & \cms                           & The hydrogen column density inferred assuming the  & Yes \\
                         &                                & power-law spectral model fitted to a custom-built  & Yes \\
                         &                                & spectrum.                                          & Yes \\
FittedPowGamma           &                                & The spectral photon index inferred assuming the    & Yes \\
                         &                                & power-law spectral model fitted to a custom-built  & Yes \\
                         &                                & spectrum.                                          & Yes \\
FittedPowFlux            & \ergcms                        & The mean 0.3--10 keV observed flux assuming the    & Yes \\
                         &                                & power-law spectral model fitted to a custom-built  & Yes \\
                         &                                & spectrum.                                          & Yes \\
FittedPowUnabsFlux       & \ergcms                        & The mean 0.3--10 keV unabsorbed flux assuming the  & Yes \\
                         &                                & power-law spectral model fitted to a custom-built  & Yes \\
                         &                                & spectrum.                                          & Yes \\
FittedPowCstat           &                                & The C-statistic from the power-law spectral fit    &  \\
                         &                                & to the custom-built spectrum.                      &  \\
FittedPowDOF             &                                & The number of degrees of freedom in the power-law  &  \\
                         &                                & spectral fit to the custom-built spectrum.         &  \\
FittedPowReducedChi2     &                                & The Churazov-weighted reduced $\chi^2$ from the     &  \\
                         &                                & power-law spectral fit to the custom-built         &  \\
                         &                                & spectrum.                                          &  \\
FittedAPECECFO           & \ecf                           & The Energy Conversion Factor from 0.3--10 keV      &  \\
                         &                                & count-rate to 0.3--10 keV observed flux, assuming  &  \\
                         &                                & the APEC spectral model fitted to a custom-built   &  \\
                         &                                & spectrum.                                          &  \\
FittedAPECECFU           & \ecf                           & The Energy Conversion Factor from 0.3--10 keV      &  \\
                         &                                & count-rate to 0.3--10 keV unabsorbed flux,         &  \\
                         &                                & assuming the APEC spectral model fitted to a       &  \\
                         &                                & custom-built spectrum.                             &  \\
FittedAPECNH             & \cms                           & The hydrogen column density inferred assuming the  & Yes \\
                         &                                & APEC spectral model fitted to a custom-built       & Yes \\
                         &                                & spectrum.                                          & Yes \\
FittedAPECkT             &  keV                           & The temperature inferred assuming the APEC         & Yes \\
                         &                                & spectral model fitted to a custom-built spectrum.  & Yes \\
FittedAPECFlux           & \ergcms                        & The mean 0.3--10 keV observed flux assuming the    & Yes \\
                         &                                & APEC spectral model fitted to a custom-built       & Yes \\
                         &                                & spectrum.                                          & Yes \\
FittedAPECUnabsFlux      & \ergcms                        & The mean 0.3--10 keV unabsorbed flux assuming the  & Yes \\
                         &                                & APEC spectral model fitted to a custom-built       & Yes \\
                         &                                & spectrum.                                          & Yes \\
FittedAPECCstat          &                                & The C-statistic from the APEC spectral fit to the  &  \\
                         &                                & custom-built spectrum.                             &  \\
FittedAPECDOF            &                                & The number of degrees of freedom in the APEC       &  \\
                         &                                & spectral fit to the custom-built spectrum.         &  \\
FittedAPECReducedChi2    &                                & The Churazov-weighted reduced $\chi^2$ from the     &  \\
                         &                                & APEC spectral fit to the custom-built spectrum.    &  \\
HasSpec                  & [bool]                         & Whether a custom-built spectrum was created for    &  \\
                         &                                & this source.                                       &  \\
\multicolumn{4}{c}{\emph{External catalogue cross-correlation results}} \\
NumExternalMatches       & [bool]                         & The number of external sources found to agree      &  \\
                         &                                & spatially with this one at the 3-\s level.         &  \\
NumExternalMatches\_slim  & [bool]                         & The number of external sources found to agree      &  \\
                         &                                & spatially with this one at the 3-\s level,         &  \\
                         &                                & excluding 2MASS, USNO-B1 and ALLWISE matches.      &  \\
MatchInROSHRI            & [bool]                         & Whether the source is spatially within 3 \s of a   &  \\
                         &                                & source in the Rosat HRI catalogue.                 &  \\
MatchIn2RXS              & [bool]                         & Whether the source is spatially within 3 \s of a   &  \\
                         &                                & source in the 2RXS catalogue.                      &  \\
MatchIn4XMM\_DR10        & [bool]                         & Whether the source is spatially within 3 \s of a   &  \\
                         &                                & source in the 3XMM-DR8 catalogue.                  &  \\
MatchIn4XMM\_DR10s       & [bool]                         & Whether the source is spatially within 3 \s of a   &  \\
                         &                                & source in the 3XMM-DR7s (stacked) catalogue.       &  \\
MatchInXMMSL2            & [bool]                         & Whether the source is spatially within 3 \s of a   &  \\
                         &                                & source in the XMMSL2 catalogue.                    &  \\
MatchInSwiftFT           & [bool]                         & Whether the source is spatially within 3 \s of a   &  \\
                         &                                & source in the SwiftFT catalogue.                   &  \\
MatchIn1SWXRT            & [bool]                         & Whether the source is spatially within 3 \s of a   &  \\
                         &                                & source in the 1SWXRT catalogue.                    &  \\
MatchInXRTGRB            & [bool]                         & Whether the source is spatially within 3 \s of an  &  \\
                         &                                & XRT GRB afterglow.                                 &  \\
MatchInSDSS\_QSO\_DR14   & [bool]                         & Whether the source is spatially within 3 \s of a   &  \\
                         &                                & source in the SDSS QSO DR14 catalogue.             &  \\
MatchIn2MASS             & [bool]                         & Whether the source is spatially within 3 \s of a   &  \\
                         &                                & source in the 2MASS catalogue.                     &  \\
MatchInUSNOB1            & [bool]                         & Whether the source is spatially within 3 \s of a   &  \\
                         &                                & source in the USNO-B1 catalogue.                   &  \\
MatchIn2CSC              & [bool]                         & Whether the source is spatially within 3 \s of a   &  \\
                         &                                & source in the 2CSC catalogue.                      &  \\
MatchIn2SXPS             & [bool]                         & Whether the source is spatially within 3 \s of a   &  \\
                         &                                & source in the 2SXPS catalogue.                     &  \\

\end{longtable}

\newpage

\subsection{Datasets}

The `Datasets' table contains information about every dataset (observations and stacked image) in LSXPS.
Observations, once in this table, will not be removed. Stacked images may be removed as a newer version
of a stack is added. If a stack is superseded, the final vertsion of it will remain in this table (see Section~\ref{sec:oldStacks}).


\begin{longtable}{cclc}
  \caption{Contents of the `Datasets' database table.}
  \label{tab:dataset} \\
  \hline
Field & Units & Description & Has Errors? \\
  \hline
  \vspace{0.1cm}
  \endfirsthead
  \multicolumn{4}{c}%
{{\bfseries \tablename\ \thetable{} -- continued from previous page}} \\
\hline
Field & Units & Description & Has Errors? \\
\hline
\vspace{0.1cm}
\endhead
\multicolumn{4}{r}{{\emph{Continued\ldots}}} \\ \hline
\endfoot
\hline
\endlastfoot

DatasetID          &                                & The unique identifier for this dataset.            &  \\
ObsID              &                                & The identifier of the observation or stacked       &  \\
                   &                                & image in which this detection occurred.            &  \\
DataVersion        &                                & The version of this dataset; for stacked images    &  \\
                   &                                & it is the number of times the image has been       &  \\
                   &                                & analysed with extra data; for single observations  &  \\
                   &                                & it is the number of times the original data were   &  \\
                   &                                & processed after data downlink.                     &  \\
IsStackedImage     & [bool]                         & Whether or not this is a stacked image.            &  \\
FieldFlag          &                                & The warning flag associated with this field.       &  \\
RA                 & Deg                            & The Right Ascension (J2000) of the field centre    &  \\
                   &                                & in decimal degrees.                                &  \\
Decl               & Deg                            & The declination (J2000) of the field centre in     &  \\
                   &                                & decimal degrees.                                   &  \\
l                  & Deg                            & Galactic longitude of the field centre.            &  \\
b                  & Deg                            & Galactic latitude of the field centre.             &  \\
ImageSize          &                                & The side length of the field image in XRT pixels   &  \\
                   &                                & (2.357 arcsec per pixel).                          &  \\
ExposureUsed       & s                              & The nominal exposure in the dataset, after all     &  \\
                   &                                & screening has been carried out.                    &  \\
StartTime\_UTC     & UTC                            & The UTC start time of the dataset.                 &  \\
StopTime\_UTC      & UTC                            & The UTC end time of the dataset.                   &  \\
OriginalExposure   & s                              & The original nominal exposure in the dataset       &  \\
                   &                                & before screening.                                  &  \\
StartTime\_MET     & MET                            & The start time of the dataset, in Swift Mission    &  \\
                   &                                & Elapsed Time.                                      &  \\
StopTime\_MET      & MET                            & The end time of the dataset, in Swift Mission      &  \\
                   &                                & Elapsed Time.                                      &  \\
MidTime\_MET       & MET                            & The mid-time of the dataset, in Swift Mission      &  \\
                   &                                & Elapsed Time.                                      &  \\
MidTime\_TDB       &                                & The mid-time of the dataset, in Barycentric        &  \\
                   &                                & Dynamical Time.                                    &  \\
MidTime\_MJD       &                                & The mid-time of the dataset, as a Modified Julian  &  \\
                   &                                & Date.                                              &  \\
LiveDate           & UTC                            & The UTC date at which this dataset became live.    &  \\
FieldBG\_band0     & ct s$^{-1}$ pixel$^{-1}$       & The mean background level in the 0.3--10 keV       &  \\
                   &                                & band.                                              &  \\
FieldBG\_band1     & ct s$^{-1}$ pixel$^{-1}$       & The mean background level in the 0.3--1 keV band.  &  \\
FieldBG\_band2     & ct s$^{-1}$ pixel$^{-1}$       & The mean background level in the 1-2 keV band.     &  \\
FieldBG\_band3     & ct s$^{-1}$ pixel$^{-1}$       & The mean background level in the 2--10 keV band.   &  \\
NumSrc\_band0      &                                & The number of sources detected in this dataset in  &  \\
                   &                                & the 0.3--10 keV band.                              &  \\
NumOK\_band0       &                                & The number of good or reasonable sources detected  &  \\
                   &                                & in this dataset in the 0.3--10 keV band.           &  \\
MedianDist\_band0  &                                & The median distance between sources detected in    &  \\
                   &                                & this dataset in the 0.3--10 keV band.              &  \\
NumSrc\_band1      &                                & The number of sources detected in this dataset in  &  \\
                   &                                & the 0.3--1 keV band.                               &  \\
NumOK\_band1       &                                & The number of good or reasonable sources detected  &  \\
                   &                                & in this dataset in the 0.3--1 keV band.            &  \\
MedianDist\_band1  &                                & The median distance between sources detected in    &  \\
                   &                                & this dataset in the 0.3--1 keV band.               &  \\
NumSrc\_band2      &                                & The number of sources detected in this dataset in  &  \\
                   &                                & the 1-2 keV band.                                  &  \\
NumOK\_band2       &                                & The number of good or reasonable sources detected  &  \\
                   &                                & in this dataset in the 1-2 keV band.               &  \\
MedianDist\_band2  &                                & The median distance between sources detected in    &  \\
                   &                                & this dataset in the 1-2 keV band.                  &  \\
NumSrc\_band3      &                                & The number of sources detected in this dataset in  &  \\
                   &                                & the 2--10 keV band.                                &  \\
NumOK\_band3       &                                & The number of good or reasonable sources detected  &  \\
                   &                                & in this dataset in the 2--10 keV band.             &  \\
MedianDist\_band3  &                                & The median distance between sources detected in    &  \\
                   &                                & this dataset in the 2--10 keV band.                &  \\
NumberOfSnapshots  &                                & The number of snapshots contributing to this       &  \\
                   &                                & dataset.                                           &  \\
AstromError        &                                & The 90\% confidence radial uncertainty on the      &  \\
                   &                                & XRT-2MASS astrometric correction.                  &  \\
CRVAL1\_corr       &                                & The CRVAL1 WCS reference value for the dataset     &  \\
                   &                                & derived from the XRT-2MASS astrometric             &  \\
                   &                                & correction.                                        &  \\
CRVAL2\_corr       &                                & The CRVAL2 WCS reference value for the dataset     &  \\
                   &                                & derived from the XRT-2MASS astrometric             &  \\
                   &                                & correction.                                        &  \\
CROTA2\_corr       &                                & The CROTA1 WCS reference value for the dataset     &  \\
                   &                                & derived from the XRT-2MASS astrometric             &  \\
                   &                                & correction.                                        &  \\
IsObsoleteStack    &  [bool]                        & Whether this corresponds to a stacked image which  &  \\
                   &                                & has been superseded by a newer version, but which  &  \\
                   &                                & contains a unique detection of a source.           &  \\

\end{longtable}

\newpage

\subsection{Detections}

The `Detections' table contains the information about every individual detection in every band, before these
are combined to create sources.


\begin{longtable}{cclc}
  \caption{Contents of the `Detections' database table.}
  \label{tab:detw} \\
  \hline
Field & Units & Description & Has Errors? \\
  \hline
  \vspace{0.1cm}
  \endfirsthead
  \multicolumn{4}{c}%
{{\bfseries \tablename\ \thetable{} -- continued from previous page}} \\
\hline
Field & Units & Description & Has Errors? \\
\hline
\vspace{0.1cm}
\endhead
\multicolumn{4}{r}{{\emph{Continued\ldots}}} \\ \hline
\endfoot
\hline
\endlastfoot

DetectionID              &                                & A unique identifier for this detection.            &  \\
DatasetID                &                                & The unique identifier for the dataset in which     &  \\
                         &                                & the detection occurred.                            &  \\
ObsID                    &                                & The obsID of the observation in which the          &  \\
                         &                                & detection occurred.                                &  \\
ObsSourceID              &                                & The ID of the obsSource this detection is part     &  \\
                         &                                & of.                                                &  \\
SourceNo                 &                                & The identifier of this source within this obsid    &  \\
                         &                                & and band.                                          &  \\
Band                     &                                & The energy band (0-3) in which this detection      &  \\
                         &                                & occurred.                                          &  \\
CorrectedExposure        & s                              & The exposure time at the position of the source    &  \\
                         &                                & in this obsID.                                     &  \\
ExposureFraction         & s                              & The fractional exposure at the position of this    &  \\
                         &                                & source, i.e. the exposure divided by the nominal   &  \\
                         &                                & exposure for the field.                            &  \\
OffaxisAngle             & arcmin                         & The angular distance of the source from the XRT    &  \\
                         &                                & boresight in all observations within which the     &  \\
                         &                                & source was detected .                              &  \\
RA                       & Deg                            & Right Ascension (J2000) in decimal degrees.        & Yes \\
Decl                     & Deg                            & Decliniation (J2000) in decimal degrees.           & Yes \\
Err90                    & arcsec                         & Position uncertainty, 90\% confidence, radial,     &  \\
                         &                                & assumed to be Rayleigh-distributed.                &  \\
RA\_corrected            & Deg                            & Right Ascension (J2000) in decimal degrees         &  \\
                         &                                & corrected using XRT-2MASS astrometry.              &  \\
Decl\_corrected          & Deg                            & Declination (J2000) in decimal degrees corrected   &  \\
                         &                                & using XRT-2MASS astrometry.                        &  \\
Err90\_corrected         & arcsec                         & Uncertainty on the corrected position, 90\%        &  \\
                         &                                & confidence, radial, assumed to be                  &  \\
                         &                                & Rayleigh-distributed.                              &  \\
l                        & Deg                            & Galactic longitude.                                &  \\
b                        & Deg                            & Galactic latitude.                                 &  \\
l\_corrected             & Deg                            & Galactic longitude corrected using XRT-2MASS       &  \\
                         &                                & astrometry.                                        &  \\
b\_corrected             & Deg                            & Galactic latitude corrected using XRT-2MASS        &  \\
                         &                                & astrometry.                                        &  \\
IMG\_X                   &                                & The x position of the object in the SKY image      &  \\
                         &                                & plane.                                             &  \\
IMG\_Y                   &                                & The y position of the object in the SKY image      &  \\
                         &                                & plane.                                             &  \\
NearestNeighbour         & arcsec                         & The distance  to the closest detection to this     &  \\
                         &                                & one, in this image.                                &  \\
NearestOKNeighbour       & arcsec                         & The distance  to the closest Good or Reasonable    &  \\
                         &                                & detection to this one, in this image.              &  \\
DetFlag                  &                                & The detection flag for this detection.             &  \\
OpticalLoadingWarning    &                                & Whether this detection is potentially affected by  &  \\
                         &                                & optical loading. If this is 0 there is no          &  \\
                         &                                & catalogue bright optical source nearby, otherwise  &  \\
                         &                                & there is a nearby optical source brighter than     &  \\
                         &                                & the level at which optical loading becomes a       &  \\
                         &                                & concern; the value indicates how many magnitudes   &  \\
                         &                                & brighter than that level the source is.            &  \\
StrayLightWarning        &                                & Whether this  detection occurred within fitted     &  \\
                         &                                & stray light rings.                                 &  \\
NearBrightSourceWarning  &                                & Whether this detection	occurred within the PSF     &  \\
                         &                                & wings of a fitted bright source.                   &  \\
MatchesKnownExtended     & [bool]                         & Whether the position of this source matches a      &  \\
                         &                                & known extended X-ray source.                       &  \\
PileupFitted             &                                & Whether the accepted fit included pile up.         &  \\
SNR                      &                                & The signal to noise ratio of the detection.        &  \\
CtsInPSFFit              &                                & Number of counts in the image region over which    &  \\
                         &                                & the final PSF fit was performed.                   &  \\
BGRateInPSFFit           & s$^{-1}$                       & Mean count-rate in the background map in the       &  \\
                         &                                & region over which the final PSF fit was            &  \\
                         &                                & performed.                                         &  \\
Cstat                    &                                & The C-statistic value from the PSF fit.            &  \\
Cstat\_nosrc             &                                & The C-statistic value derived over the PSF fit     &  \\
                         &                                & region if no source is fitted.                     &  \\
L\_src                   &                                & The log-likelihood value that this detection is    &  \\
                         &                                & not just a background fluctuation.                 &  \\
Cstat\_flat              &                                & The Cstat value found from a fit assuming a        &  \\
                         &                                & constant increase above the background (i.e. the   &  \\
                         &                                & count excess is flat, not PSF-like).               &  \\
Lflat                    &                                & The log-likelihood value that this detection is    &  \\
                         &                                & PSF like, not a flat count increase.               &  \\
FracPix                  &                                & The fraction of pixels within the PSF fit region   &  \\
                         &                                & which are exposed.                                 &  \\
Pileup\_S                &                                & The best-fitting S parameter of the pile-up        &  \\
                         &                                & model, if fitted.                                  &  \\
Pileup\_l                &                                & The best-fitting l parameter of the pile-up        &  \\
                         &                                & model, if fitted.                                  &  \\
Pileup\_c                &                                & The best-fitting c parameter of the pile-up        &  \\
                         &                                & model, if fitted.                                  &  \\
Pileup\_tau              &                                & The best-fittingtau parameter of the pile-up       &  \\
                         &                                & model, if fitted.                                  &  \\
Cstat\_altPileup         &                                & The Cstat value found from the unusued fit. i.e.   &  \\
                         &                                & if the piled up model was used, this gives the     &  \\
                         &                                & Cstat from the non-piled-up fit, and vice-versa.   &  \\
PSF\_Fit\_Radius         &                                & The radius of the circular region over which PSF   &  \\
                         &                                & fitting was carried out (in image pixels,          &  \\
                         &                                & 2.357\arcsec\ to a side).                           &  \\
CellDetect\_BoxWidth     &                                & The full width of the cell-detect box in which     &  \\
                         &                                & this source was detected (in image pixels,         &  \\
                         &                                & 2.357 \arcsec\ to a side).                           &  \\
Rate                     & s$^{-1}$                       & The corrected count-rate of this detection (i.e.   & Yes \\
                         &                                & in the band of this image).                        & Yes \\
CtsInRate                & s$^{-1}$                       & The total number of counts in the region used to   &  \\
                         &                                & extract the count rate.                            &  \\
BGCtsInRate              & s$^{-1}$                       & The total number of counts in the region used to   &  \\
                         &                                & extract the count rate.                            &  \\
Rate\_CF                 & s$^{-1}$                       & The PSF correction factor for the count rate.      &  \\
BGRateInRate             & s$^{-1}$                       & The background rate in the region used for         &  \\
                         &                                & count-rate calculation.                            &  \\

\end{longtable}

\newpage

\subsection{ObsSources}

As described in Appendix~\ref{sec:appBuild}, the process of identifying unique sources has two steps.
In the first, individual detections from different bands within a dataset are merged to identify the objects
uniquely detected in that dataset. These are referred to as \emph{ObsSources} and are available in this catalogue table.


\begin{longtable}{cclc}
  \caption{Contents of the `ObsSource' database table.}
  \label{tab:os} \\
  \hline
Field & Units & Description & Has Errors? \\
  \hline
  \vspace{0.1cm}
  \endfirsthead
  \multicolumn{4}{c}%
{{\bfseries \tablename\ \thetable{} -- continued from previous page}} \\
\hline
Field & Units & Description & Has Errors? \\
\hline
\vspace{0.1cm}
\endhead
\multicolumn{4}{r}{{\emph{Continued\ldots}}} \\ \hline
\endfoot
\hline
\endlastfoot

ObsSourceID            &                                & The unique identifier of this ObsSource.           &  \\
DatasetID              &                                & The dataset identifier in which this obsSource     &  \\
                       &                                & was found.                                         &  \\
LSXPS\_ID              &                                & Numerical unique source identifier within LSXPS.   &  \\
OSNum                  &                                & The identifier of this obsSource within this       &  \\
                       &                                & dataset.                                           &  \\
UsedCorrectedPosition  & [bool]                         & Whether or not the astrometrically-corrected       &  \\
                       &                                & position of this obsSource should be used.         &  \\
NearestNeighbour       & arcsec                         & The distance  to the closest LSXPS obsSource       &  \\
                       &                                & within this dataset.                               &  \\
NearestOKNeighbour     & arcsec                         & The distance  to the closest good or reasonable    &  \\
                       &                                & LSXPS obsSource within this dataset.               &  \\
Exposure               & s                              & The exposure time at the position of the           &  \\
                       &                                & obsSource, corrected for vignetting.               &  \\
HR1                    &                                & The aggregate HR1 hardness ratio of the            & Yes \\
                       &                                & obsSource.                                         & Yes \\
HR2                    &                                & The aggregate HR2 hardness ratio of the            & Yes \\
                       &                                & obsSource.                                         & Yes \\
BestBand               &                                & The energy band corresponding to the best          &  \\
                       &                                & detection of this obsSource in this dataset.       &  \\
PileupWarning          & [bool]                         & Whether a pile up warning exists, indicating that  &  \\
                       &                                & the object is very piled up and in some bands      &  \\
                       &                                & there is evidence of pile up not being fitted.     &  \\
BestDetectionID        &                                & The DetectionID of the best detection of this      &  \\
                       &                                & obsSource in this dataset.                         &  \\
IsOrphanSource         & [bool]                         & Whether this source is one which is only detected  &  \\
                       &                                & in this dataset, and this dataset is deprecated.   &  \\

\end{longtable}

\newpage

\subsection{External catalogue matches}

As with previous SXPS catalogues, we check each source for possible counterparts in other catalogues.
Details of those matches are given in this table, which is entitled `xCorr' in the file names, for brevity.


\begin{longtable}{cclc}
  \caption{Contents of the `xCorr' database table.}
  \label{tab:xcorr} \\
  \hline
Field & Units & Description & Has Errors? \\
  \hline
  \vspace{0.1cm}
  \endfirsthead
  \multicolumn{4}{c}%
{{\bfseries \tablename\ \thetable{} -- continued from previous page}} \\
\hline
Field & Units & Description & Has Errors? \\
\hline
\vspace{0.1cm}
\endhead
\multicolumn{4}{r}{{\emph{Continued\ldots}}} \\ \hline
\endfoot
\hline
\endlastfoot

LSXPS\_ID  &                                & The numerical identifier of the LSXPS source.      &  \\
ExtCat\_ID  &                                & The catalogued name of the external source.        &  \\
Catalogue  &                                & The catalogue in which the external source was     &  \\
           &                                & found.                                             &  \\
Distance   & arcsec                         & The distance in arcseconds between the 2SXPS and   &  \\
           &                                & external source positions.                         &  \\
Ext\_RA    & Deg                            & The Right Ascension (J2000) in decimal degrees of  &  \\
           &                                & the external source.                               &  \\
Ext\_Decl  & Deg                            & The declination (J2000) in decimal degrees of the  &  \\
           &                                & external source.                                   &  \\
Ext\_Err90  & arcsec                         & The 90\% confidence radial position error of the   &  \\
           &                                & external source used for cross correlation. This   &  \\
           &                                & may have been converted to 90\% (assuming          &  \\
           &                                & Rayleigh statistics) and had a systematic error    &  \\
           &                                & added if necessary, therefore may differ from the  &  \\
           &                                & catalogued value.                                  &  \\
\end{longtable}

\subsection{OldStacks}

When a new version of a stacked image is processed, or the stack superseded (see Section~\ref{sec:oldStacks}),
some brief information about the old version is stored in this table.


\begin{longtable}{cclc}
  \caption{Contents of the `oldStacks' database table.}
  \label{tab:oldstack} \\
  \hline
Field & Units & Description & Has Errors? \\
  \hline
  \vspace{0.1cm}
  \endfirsthead
  \multicolumn{4}{c}%
{{\bfseries \tablename\ \thetable{} -- continued from previous page}} \\
\hline
Field & Units & Description & Has Errors? \\
\hline
\vspace{0.1cm}
\endhead
\multicolumn{4}{r}{{\emph{Continued\ldots}}} \\ \hline
\endfoot
\hline
\endlastfoot

DatasetID        &                                & The unique identifier for this dataset.            &  \\
DataVersion      &                                & The version of this dataset; for stacked images    &  \\
                 &                                & it is the number of times the image has been       &  \\
                 &                                & analysed with extra data; for single observations  &  \\
                 &                                & it is the number of times the original data were   &  \\
                 &                                & processed after data downlink.                     &  \\
LiveDate         & UTC                            & The UTC date at which this dataset became live.    &  \\
ObsID            &                                & The identifier of the observation or stacked       &  \\
                 &                                & image in which this detection occurred.            &  \\
IsObsoleteStack  & [bool]                         & Whether this corresponds to a stacked image which  &  \\
                 &                                & has been superseded by a newer version, but which  &  \\
                 &                                & contains a unique detection of a source.           &  \\

\end{longtable}

\newpage

\subsection{Transients}

The `Transients' table contains the information about every transient counterpart that
has been checked by a member of the XRT team and classed into one of the `possible transient'
categories (Section~\ref{sec:transCheck}).


\begin{longtable}{cclc}
  \caption{Contents of the `Transients' database table.}
  \label{tab:transtab} \\
  \hline
Field & Units & Description & Has Errors? \\
  \hline
  \vspace{0.1cm}
  \endfirsthead
  \multicolumn{4}{c}%
{{\bfseries \tablename\ \thetable{} -- continued from previous page}} \\
\hline
Field & Units & Description & Has Errors? \\
\hline
\vspace{0.1cm}
\endhead
\multicolumn{4}{r}{{\emph{Continued\ldots}}} \\ \hline
\endfoot
\hline
\endlastfoot

TransientID                                     &                                & A unique numerical identifier for this transient   &  \\
                                                &                                & within this catalogue.                             &  \\
IAUName                                         &                                & The IAU-format name of the transient: SwiftJ       &  \\
                                                &                                & HHMMSS.S+ddmmss.                                   &  \\
LSXPS\_ID                                       &                                & The source identifier within the main LSXPS        &  \\
                                                &                                & catalogue.                                         &  \\
LSXPSName                                       &                                & The IAU name of this object in the main LSXPS      &  \\
                                                &                                & catalogue.                                         &  \\
Classification                                  &                                & A numerical code describing how this transient     &  \\
                                                &                                & has been classified.                               &  \\
RA                                              & Deg                            & Right Ascension (J2000) in decminal degrees.       &  \\
Decl                                            & Deg                            & Declination (J2000) in decimal degrees.            &  \\
Err90                                           & arcsec                         & Position uncertainty, 90\% confidence, radial,     &  \\
                                                &                                & assumed to be Rayleigh-distributed.                &  \\
l                                               & Deg                            & Galactic longitude.                                &  \\
b                                               & Deg                            & Galactic latitude.                                 &  \\
DiscoveryDatasetID                              &                                & The date of the observation in which the           &  \\
                                                &                                & transient was first found.                         &  \\
ObsSourceID                                     &                                & The numerical identified of the first detection    &  \\
                                                &                                & in the Public\_ObsSources table.                   &  \\
PeakRateAtDetection                             & s$^{-1}$                       & The peak count-rate in the 0.3--10 keV band in     & Yes \\
                                                &                                & the dataset in which the transient was first       & Yes \\
                                                &                                & detected.                                          & Yes \\
PeakSoftMedBandRateAtDetection                  & s$^{-1}$                       & The peak count-rate in the 0.3--2 keV band in the  &  \\
                                                &                                & dataset in which the transient was first           &  \\
                                                &                                & detected.                                          &  \\
UpperLimitSource\_Canned                        &                                & Where the `best' upper limit at the source     &  \\
                                                &                                & location was taken from, where the `canned'    &  \\
                                                &                                & spectrum is used to convert between                &  \\
                                                &                                & observatories.                                     &  \\
UpperLimitSource\_DiscoverySpectrum             &                                & Where the `best' upper limit at the source     &  \\
                                                &                                & location was taken from, where the discovery       &  \\
                                                &                                & spectrum is used to convert between                &  \\
                                                &                                & observatories.                                     &  \\
UpperLimitSource\_FullSpectrum                  &                                & Where the `best' upper limit at the source     &  \\
                                                &                                & location was taken from, where the full spectrum   &  \\
                                                &                                & is used to convert between observatories.          &  \\
Significance                                    &                                & How far above the best upper limit the peak count  &  \\
                                                &                                & rate is, in Gaussian \ss.                          &  \\
InitSpecNH                                      & \cms                           & The hydrogen column density found in an automated  &  \\
                                                &                                & power-law fit to a spectrum built from the         &  \\
                                                &                                & discovery dataset.                                 &  \\
InitSpecGamma                                   &                                & The photon index found in an automated power-law   &  \\
                                                &                                & fit to a spectrum built from the discovery         &  \\
                                                &                                & dataset.                                           &  \\
InitSpecCstat                                   &                                & The C-statistic from an automated power-law fit    &  \\
                                                &                                & to a spectrum built from the discovery dataset.    &  \\
InitSpecTestStat                                &                                & The  Churazov-weighted reduced $\chi^2$ from an     &  \\
                                                &                                & automated power-law fit to a spectrum built from   &  \\
                                                &                                & the discovery dataset.                             &  \\
InitSpecDof                                     &                                & The number of degrees of freedom in the automated  &  \\
                                                &                                & power-law fit to a spectrum built from the         &  \\
                                                &                                & discovery dataset.                                 &  \\
FullSpecNH                                      & \cms                           & The hydrogen column density found in an automated  &  \\
                                                &                                & power-law fit to a spectrum built from all         &  \\
                                                &                                & observations of the transient after its            &  \\
                                                &                                & discovery.                                         &  \\
FullSpecGamma                                   &                                & The photon index found in an automated power-law   &  \\
                                                &                                & fit to a spectrum built from all observations of   &  \\
                                                &                                & the transient after its discovery.                 &  \\
FullSpecCstat                                   &                                & The C-statistic from an automated power-law fit    &  \\
                                                &                                & to a spectrum built from all observations of the   &  \\
                                                &                                & transient after its discovery.                     &  \\
FullSpecTestStat                                &                                & The  Churazov-weighted reduced $\chi^2$ from an     &  \\
                                                &                                & automated power-law fit to a spectrum built from   &  \\
                                                &                                & all observations of the transient after its        &  \\
                                                &                                & discovery.                                         &  \\
FullSpecDof                                     &                                & The number of degrees of freedom in the automated  &  \\
                                                &                                & power-law fit to a spectrum built from the all     &  \\
                                                &                                & observations of the transient after its            &  \\
                                                &                                & discovery.                                         &  \\
DiscoveryDate                                   & UTC                            & The date of the analysis in which the transient    &  \\
                                                &                                & was first found.                                   &  \\
DetectionDate                                   & UTC                            & The MET of the observation in which the transient  &  \\
                                                &                                & was first found.                                   &  \\
DetectionMET                                    & MET                            & The date of the observation in which the           &  \\
                                                &                                & transient was first found.                         &  \\
LSXPS\_UpperLimit                               & s$^{-1}$                       & An upper limit at the transient derived from the   &  \\
                                                &                                & LSXPS catalogue.                                   &  \\
LSXPS\_UpperLimit\_ObsID                        &                                & The observation in the LSXPS catalogue from which  &  \\
                                                &                                & the upper limit was derived.                       &  \\
XMM\_UpperLimit\_native                         & s$^{-1}$                       & An XMM-Newton upper limit at the transient         &  \\
                                                &                                & location, obtained via the HILIGT server, in       &  \\
                                                &                                & native XMM count-rate units.                       &  \\
XMM\_UpperLimit\_ObsMode                        &                                & The mode in which XMM was observing when the       &  \\
                                                &                                & upper limit was obtained.                          &  \\
XMM\_UpperLimit\_Instrument                     &                                & The XMM-Newton instrument from which the upper     &  \\
                                                &                                & limit was obtained.                                &  \\
XMM\_UpperLimit\_asXRTTotal\_DiscoverySpectrum  &  s$^{-1}$                      & The XMM-Newton upper limit converted to a 0.3--10  &  \\
                                                &                                & keV XRT count rate, using the automated power-law  &  \\
                                                &                                & fit to a spectrum built from the discovery         &  \\
                                                &                                & dataset.                                           &  \\
XMM\_UpperLimit\_asXRTTotal\_FullSpectrum       &  s$^{-1}$                      & The XMM-Newton upper limit converted to a 0.3--10  &  \\
                                                &                                & keV XRT count rate, using the automated power-law  &  \\
                                                &                                & fit to a spectrum built from the all observations  &  \\
                                                &                                & of the transient after its discovery.              &  \\
XMM\_UpperLimit\_asXRTTotal\_Canned             &  s$^{-1}$                      & The XMM-Newton upper limit converted to a 0.3--10  &  \\
                                                &                                & keV XRT count rate, using a standard AGN           &  \\
                                                &                                & spectrum.                                          &  \\
RASS\_UpperLimit\_native                        &  s$^{-1}$                      & An ROSAT All-Sky Survey upper limit at the         &  \\
                                                &                                & transient location, obtained via the HILIGT        &  \\
                                                &                                & server, in native ROSAT PSPC count-rate units.     &  \\
RASS\_UpperLimit\_asXRTSoftMed\_DiscoverySpectrum  & s$^{-1}$                       & The RASS upper limit converted to a 0.3--2 keV     &  \\
                                                &                                & XRT count rate, using the automated power-law fit  &  \\
                                                &                                & to a spectrum built from the discovery dataset.    &  \\
RASS\_UpperLimit\_asXRTSoftMed\_FullSpectrum    &  s$^{-1}$                      & The RASS upper limit converted to a 0.3--2 keV     &  \\
                                                &                                & XRT count rate, using the automated power-law fit  &  \\
                                                &                                & to a spectrum built from the all observations of   &  \\
                                                &                                & the transient after its discovery.                 &  \\
RASS\_UpperLimit\_asXRTSoftMed\_Canned          &  s$^{-1}$                      & The RASS upper limit converted to a 0.3--2 keV     &  \\
                                                &                                & XRT count rate, using a standard AGN spectrum.     &  \\
RASS\_UpperLimit\_asXRTTotal\_DiscoverySpectrum  &  s$^{-1}$                     & The RASS upper limit converted to a 0.3--10 keV    &  \\
                                                &                                & XRT count rate, using the automated power-law fit  &  \\
                                                &                                & to a spectrum built from the discovery dataset.    &  \\
RASS\_UpperLimit\_asXRTTotal\_FullSpectrum      &  s$^{-1}$                      & The RASS upper limit converted to a 0.3--10 keV    &  \\
                                                &                                & XRT count rate, using the automated power-law fit  &  \\
                                                &                                & to a spectrum built from the all observations of   &  \\
                                                &                                & the transient after its discovery.                 &  \\
RASS\_UpperLimit\_asXRTTotal\_Canned            &   s$^{-1}$                     & The RASS upper limit converted to a 0.3--10 keV    &  \\
                                                &                                & XRT count rate, using a standard AGN spectrum.     &  \\
Notes                                           &                                & Any notes added by the XRT team.                   &  \\

\end{longtable}

\label{lastpage}

\end{document}